\documentclass[prx,twocolumn,floatfix,superscriptaddress,longbibliography]{revtex4-1}

\usepackage{amssymb,amsmath,amstext}                
\usepackage{graphicx}                                               
\usepackage{epstopdf}                                               
\usepackage{color}       
\usepackage{bm}
\usepackage{appendix}                                              
\usepackage[utf8]{inputenc}
\usepackage{bbold}
\usepackage{bbm}
\usepackage{latexsym}
\usepackage[colorlinks=true,citecolor=blue,linkcolor=magenta]{hyperref}

\begin{document}
 \title{Precise extrapolation of the correlation function asymptotics in uniform tensor network states with application to the Bose-Hubbard and XXZ models}

\author{Marek M. Rams}
\affiliation{Jagiellonian University,  Marian Smoluchowski Institute of Physics,  \\ \L{}ojasiewicza 11, PL-30348 Krak\'ow, Poland }
\author{Piotr Czarnik}
\affiliation{Institute of Nuclear Physics, Polish Academy of Sciences, Radzikowskiego 152, PL-31342 Krak\'ow, Poland}
\author{Lukasz Cincio}
\affiliation{Theory Division, Los Alamos National Laboratory, Los Alamos, NM 87545, USA}

\begin{abstract}
We analyze the problem of extracting the correlation length from infinite matrix product states (MPS) and corner transfer matrix (CTM) simulations.  When the correlation length is calculated directly from the transfer matrix, it is typically significantly underestimated for finite bond dimensions used in numerical simulation. This is true even when one considers ground states at a distance from the critical point. In this article we introduce extrapolation procedure to overcome this problem. To that end we quantify how much the dominant part of the MPS/CTM transfer matrix spectrum deviates from being continuous. The latter is necessary to capture the exact asymptotics of the correlation function where the exponential decay is typically modified by an additional algebraic term. By extrapolating such a refinement parameter to zero, we show that we are able to recover the exact value of the correlation length with high accuracy. In a generic setting, our method reduces the error by a factor of $\sim 100$ as compared to the results obtained without extrapolation and a factor of $\sim 10$ as compared to simple extrapolation schemes employing bond dimension.   We test our approach in a number of solvable models both in 1d and 2d.  Subsequently, we apply it to one-dimensional XXZ spin-$\frac32$ and the Bose-Hubbard models in a massive regime in the vicinity of the Berezinskii–Kosterlitz–Thouless critical point. We then fit the scaling form of the correlation length and extract the position of the critical point and obtain  results comparable or better than those of other state-of-the-art numerical methods. Finally, we show how the algebraic part of the correlation function asymptotics can be directly recovered from the scaling of the dominant form factor within our approach. Our method provides the means for detailed studies of phase diagrams of quantum models in 1d and, through the finite correlation length scaling of projected entangled pair states, also in 2d.
\end{abstract}
\maketitle


\section{Introduction} 
Tensor networks and related numerical renormalization group techniques allow to efficiently approximate systems of exponentially many degrees of freedom with manageable number of a few relevant ones, providing invaluable tools in the studies of strongly correlated systems. We are particularly interested in two such techniques. The first one is based on the matrix product state (MPS) representation of a many-body wave function \cite{Fannes_MPS_1992}. It provides the underlying framework behind a family of state-of-the-art methods for approximating the low-energy states of local one-dimensional Hamiltonians \cite{verstraete_review_2008, schollwock_review_2011, orus_review_2014}, descendants of the seminal density matrix renormalization group (DMRG) algorithm \cite{White_DMRG_1992, White_DMRG_1993}. The second one, closely related, is the corner transfer matrix (CTM) algorithm \cite{Nishino_CTM_1996, Nishino_CTM_1997}. It is used to numerically solve classical systems in 2d and is a method of choice for contracting 2d quantum states described by the projected entangled pair states (PEPS) ansatz \cite{Verstraete_PEPS_2004, Jordan_IPEPS_2008}.

In this article we address the problem of precise extrapolation of the correlation length in such simulations. Correlation length is a fundamental quantity in the description of (quantum) many-body systems and their phase transitions. It provides valuable input into the nature of the phase being described, as well as informs about its boundaries. Apart from this general consideration, there are two immediate applications of the presented method that we mention below. Firstly, it is desirable to establish a reliable method to calculate the correlation length of PEPS using the CTM algorithm. It can be applied to characterize the critical behavior of 2d quantum systems within the PEPS approach and set up finite correlation length scaling both at zero \cite{Rader_FCLS_2018, Corboz_FCLS_2018} and at finite temperature \cite{Czarnik_2018}. A similar strategy can also be used in studies of classical systems in 3d \cite{Vanderstraeten2018}.
Secondly, it allows to fit the scaling form of the correlation length in the vicinity of the  Berezinskii–Kosterlitz–Thouless (BKT) critical point in one-dimensional Bose-Hubbard and XXZ type models. Besides precisely extracting the critical point position, the accurate knowledge of the scaling form is relevant, for instance, to understand the behavior of quantum fidelity in such systems \cite{Cincio_fid_BH}. 

Below we focus the discussion on MPS and notice that the argument could also be directly applied to CTM. It is well known that MPS with finite bond dimension is able to reproduce the ground state of the local gapped Hamiltonian up to an error which vanishes exponentially with the bond dimension \cite{verstraete_faithfully_2006}. Consequently, local observables can usually be simulated with similar precision. The correlation length, on the other hand, describes the tail of the correlation function. This tail is vanishing exponentially and as such there is no reason to expect that MPS would be able to capture it faithfully. More importantly, asymptotics of the connected correlation function calculated for MPS with finite bond dimension is purely exponential \cite{Fannes_MPS_1992}
\begin{equation}
C(R) \sim e^{-R/\xi_{A}},
\label{eq:C_MPS}
\end{equation}
while typically one expects that the exponential decay is modified by an additional algebraic term, 
\begin{equation}
C(R) \sim R^{-\eta} e^{-R/\xi},
\label{eq:C_asymptotic}
\end{equation}
as, e.g., in the Ornstein-Zernike formula for the correlation function in the context of the Ising-type models \cite{Kennedy_OZ_1991,Campanino_OZ_2003}. For those reasons, it is not straightforward to faithfully recover the asymptotics of the correlation function directly from MPS simulations with finite bond dimension. To highlight the problem, for the specific point in the XXZ spin-$\frac12$ model which we discuss in detail later, MPS with bond dimension 4096 recovers the exact ground state energy with an error of the order of $10^{-12}$, but at the same time it is still underestimating the correlation length by a factor of 2.

In this article we propose an extrapolation scheme to overcome the above problem. For uniform MPS (CTM) which describes a translationally invariant system the correlation length is calculated from the ratio of the two largest eigenvalues of the site-to-site (column-to-column) transfer matrix. The spectrum of this transfer matrix is necessarily discrete for the finite bond dimension used in the numerical simulations. In order to recover the algebraic part of the correlation function asymptotics in Eq.~\eqref{eq:C_asymptotic}, the spectrum would have to be continuous. In our approach, we look at the distance $\delta$ between the next dominant transfer matrix eigenvalues, e.g. the second and the third one, and employ it as a measure of deviation from the exact solution. One expects to recover $\delta = 0$ in the limit of the exact representation of the ground state. For a given model, we calculate the MPS correlation length, as well as the refinement parameter $\delta$, for a number of MPSs with increasing bond dimensions and subsequently extrapolate $\delta \to 0$ in order to recover the actual value of the correlation length. In order to benchmark our approach we analyze a number of models where the correlation length, or some related properties like the position of the critical point and its universality class, are known analytically. Based on this data we argue that the method proposed in this article is more reliable and produces more accurate results than the one that directly uses the bond dimension as a refinement parameter. 

Correlation function asymptotics can be derived from the Euclidian path-integral representation of the ground state and the exact quantum transfer matrix (QTM). In Refs.~\onlinecite{Zauner_2015,Rams_2015, Bal_2016} it was argued that the MPS transfer matrix can be understood as an approximation of the QTM obtained as a result of the renormalization procedure akin to Wilson's numerical renormalization group description of the impurity problem \cite{Wilson_NRG_1975}. In this picture, the physical spin is interpreted as an impurity in -- by construction translationally invariant -- QTM, and the MPS transfer matrix retains only the degrees of freedom (along the virtual, imaginary time direction of the system) relevant for the description of correlations of such an impurity. In this article we further build on this intuition and observe how the form factors (i.e., matrix elements of the operator transfer matrix in a suitable basis, defined below) are effectively being renormalized. Most importantly, we argue that the exponent $\eta$ of the algebraic part of the correlation function asymptotics in Eq.~\eqref{eq:C_asymptotic} is directly related to how the relevant dominant form factor decays as we approach the exact solution, $\delta\to 0$ (in a limit of infinite bond dimension). 

We should finally contrast our approach with the finite entanglement scaling scheme where, for the 1d system in the vicinity of the critical point, finite MPS correlation length (a result of finite MPS bond dimension) is used similarly to the effective finite size of the system in order to postulate a scaling hypothesis. The position of the critical point and critical exponents can then be extracted by proper renormalization and collapse of the data  obtained for different bond dimensions \cite{Nishino_FES_1996, Tagliacozzo_FES_2008, Pollmann_FES_2009, Pirvu_FES_2012, Kjall_FES_2013, Stojevic_FES_2015, Ueda_FES_2017}. Our method enables obtaining the scaling form of the correlation length without assuming the scaling hypothesis and as such can be used to independently corroborate some of the results found with finite entanglement scaling. We remark that obtaining the correlation length outside the critical regimes is beyond standard finite entanglement scaling schemes as it requires knowledge of a nonuniversal scaling function.  Furthermore, our extrapolations  can also be applied beyond the scaling regime, or more generally, when it might be impossible to postulate a scaling ansatz. In this context, for instance, it should be possible to apply our method to obtain a more precise description of nonequilibrium dynamical properties, such as spreading of the correlations in the excited system. The two methods are equivalent exactly at the critical point as we note that our refinement parameter $\delta$ is directly proportional to the inverse of the MPS correlation length in this case.

More importantly, both methods have to be combined in studies of quantum systems in 2d. In that case both the limited bond dimension of PEPS and  the finite bond dimension of CTM used to contract it contribute to errors. Setting up finite correlation length scaling \cite{Rader_FCLS_2018, Corboz_FCLS_2018, Czarnik_2018} (or finite entanglement scaling, as it is often called in the context of 1d systems) requires 
 extrapolating the error of the correlation length resulting from CTM to zero in a controlled way, which is the goal of this article.  As such, we anticipate mapping out phase diagrams of quantum systems both in 1d and 2d as an important application of our method. 
 
The rest of the article is organized as follows. In Sec. \ref{sec1} we introduce relevant notation focusing on MPS and quantify the general arguments from the Introduction.  In Sec.~\ref{sec2} we briefly summarize our method discussing, in particular, how different properties of a studied model can be used to further refine the measure of error used for extrapolation. We benchmark our approach in Sec.~\ref{sec3}.  We study XY and XXZ spin-$\frac12$ models, where the exact value of the correlation length is known. Then we focus on XXZ spin-$\frac32$ model and conclude with the Bose-Hubbard model with unit filling as a nontrivial implementation of our method. In Sec.~\ref{sec4} we test our approach in the context of 2d models and their PEPS description. We employ corner transfer matrix method to analyze exactly solvable statistical models: the classical 2d Ising model and 8-vertex model. Finally, we apply our technique to thermal quantum states in 2d, including interacting spinless fermions, where the correlation length is not known otherwise. In Sec.~\ref{sec5}, we discuss how the exponent of the leading algebraic part of the correlation function naturally emerges within our approach and uncover its connection with the form factors. We conclude in Sec.~\ref{sec6}. Appendix \ref{sec:app} illustrates the problems related to fitting the asymptotic directly, where inaccurate results are obtained away from the critical point. Paradoxically, it is possible to obtain much better results for the critical systems within such an approach. This approach is indeed widely used in the literature. Finally, in Appendix~\ref{sec:app2} we argue that methods that provide good extrapolation of energy per lattice site are not suited to work well for the correlation length.
  

\section{Notation} 
 \label{sec1}
 
 In this article we focus on infinite, translationally invariant systems. Setting up the notation, uniform matrix product states take the form
\begin{equation}
|\Psi\rangle =\sum_{ \bf s } \left( \prod_{n \in \mathbb{Z}} A^{s_n} \right) |  {\bf s} \rangle,
\label{eq:uMPS}
\end{equation}
where $| {\bf s} \rangle = |\ldots, s_1,s_2,s_3,s_4,\ldots \rangle$  and  $A^{s_n}$ are $D\times D$ matrices with parameter $D$ usually referred to as the MPS bond dimension.

The MPS transfer matrix (TM) is defined in a standard way
\begin{equation}
\mathcal{T}_A  = \sum_{s=1}^d \bar{A}^{s} \otimes A^s \ .
\label{eq:TM}
\end{equation}
It is the key object in the calculation of the static correlation function. In order to calculate the expectation values related to some operator $ o$ it is also convenient to define the operator transfer matrix
\begin{equation}
\mathcal{T}_{A}^{ o} = \sum_{s,r=1}^d  o_{s,r} \bar{A}^{s} \otimes A^r.
\label{eq:oTM}
\end{equation}
For larger unit cell consisting of $L$ sites, i.e. if MPS is translationally invariant only when shifted by $L$ lattice sites (i.e. due to spontaneous breaking of translational symmetry), those $L$ sites are combined into one to calculate the transfer matrix in Eqs.~(\ref{eq:TM}, \ref{eq:oTM}).

In our approach we focus on the eigenvalues of the transfer matrix $\mathcal{T}_A$,
\begin{equation}
\lambda_j = e^{-(\epsilon_j + i \phi_j) L},
\label{eq:lambdaj}
\end{equation}
with $j=0,1,\ldots, D^2-1$ and $|\lambda_0| > |\lambda_1| \ge |\lambda_2| \ge \ldots$. Eq.~\eqref{eq:lambdaj} singles out (minus log of) the absolute value and the phase into $\epsilon_j \ge 0$ and $\phi_j \in (-\pi/L,\pi/L]$, respectively. We have introduced the period $L$, so that the correlation length is measured in the units of lattice spacing. Note that some information about phase is lost for $L>1$. Proper normalization of the state $|\Psi\rangle$ entails that $\lambda_0 = 1$. We additionally assume that the largest eigenvalue of the TM is unique. This ensures the state $| \Psi \rangle$ in Eq.~(\ref{eq:uMPS}) is properly defined.

It is well known that the correlation length $\xi_A$ associated with given normalized MPS is set by the second largest transfer matrix eigenvalue as 
\begin{equation}
\xi_A = 1/\epsilon_1,
\end{equation}
and $\phi_1$ captures the leading period of oscillations of the correlation function.  More precisely, the connected correlation function of operators $ o$ and $ q$ at distance $R$ can be expressed as
\begin{equation} 
C_{ o  q}(R)= \langle  o_0  q_R \rangle - \langle  o_0 \rangle \langle  q_R \rangle =  \sum_{j>0} f_j^{oq} e^{ -(\epsilon_j + i \phi_j) R},
\label{eq:C_TM}
\end{equation}
where the form factors $f_j^{oq}$ are defined as 
\begin{equation}
f_j^{oq} = ( \varphi_0 \vert \mathcal{T}_A^{ o}  \vert \varphi_j) ( \varphi_j \vert \mathcal{T}_A^{ q}  \vert \varphi_0 ),
\end{equation}
 with right $\vert \varphi_j )$ and left $( \varphi_j \vert$ eigenvectors of the transfer matrix normalized as $( \varphi_i \vert \varphi_j ) = \delta_{ij}$. 
Eq.~\eqref{eq:C_TM} leads to purely exponential decay of asymptotics of the correlation function as in Eq.~\eqref{eq:C_MPS}, dictated typically by the second largest eigenvalue of $\mathcal{T}_A$. More generally, this decay is established by the largest eigenvalues ($j>0$) for which the corresponding form factors $f_j^{oq}$ are nonzero.

The asymptotics of the correlation function, however, typically contains an algebraic factor as in Eq.~\eqref{eq:C_asymptotic}. For the MPS to recover the algebraic part $R^{-\eta}$ of the asymptotics, the spectrum of the transfer matrix would have to be continuous. This is clearly impossible for the finite bond dimension used in numerical simulations. This is a well-known fact in the studies of critical points. Here, the decay of the correlation function is purely algebraic. This fact is used as one of the arguments of why simulating such points with MPS is particularly challenging.  As argued above, those issues are still present also far away from the critical point in the context of precise extrapolation of the correlation length.

\begin{figure} [t]
\begin{center}
  \includegraphics[width= 0.85\columnwidth]{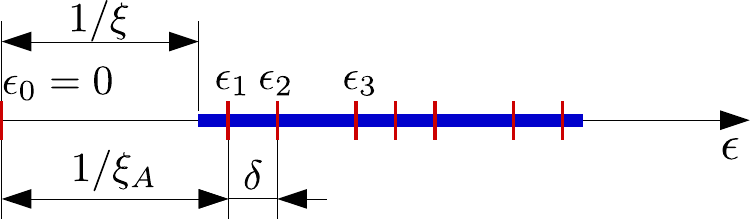}
\end{center}
  \caption{ Illustration of the idea behind the scheme. The figure represents (the logarithm of) the dominant part of the transfer matrix spectrum in a generic situation.
   The blue line represents continuous band necessary to recover the algebraic part of the correlation function asymptotics in Eq.~\eqref{eq:C_asymptotic}. In this case the exact correlation length is set by the gap between the bottom of the band and the origin, $\epsilon_0=0$. The spectrum of the transfer matrix for finite bond dimension MPS, represented here by red marks, is necessarily discrete and as such can only approximate the continuous band. Consequently, $1/\epsilon_1$ is typically underestimating the true value of the correlation length. We employ $\delta = \epsilon_2 - \epsilon_1$ as a natural measure of how well the discrete spectrum is able to approximate the exact continuous one.  By computing $\epsilon_1(D)$ and $\delta(D)$ for some number of MPSs with different bond dimensions $D$, we extract the correlation length by extrapolating $\delta \to 0$.
  }  
   \label{fig:1}
\end{figure}

The discrete dominant eigenvalues of the transfer matrix can at best approximate the continuous spectrum related to the exact quantum transfer matrix, as presented pictorially in Fig.~\ref{fig:1}.  As suggested in that picture, it can be expected that the correlation length obtained as $\xi_A = 1/\epsilon_1$ underestimates the exact value, as $\epsilon_1$ would be localized inside the band and not on its edge. Consequently,  one has to resort to extrapolation in order to recover the true value of the correlation length. 

An alternative approach would be to fit the asymptotics in Eq.~\eqref{eq:C_asymptotic} for the intermediate values of the distance $R$. However, away from the critical point, this would require the ability to fit the correlation function asymptotics for distances between
the physical correlation length and the length scale that results from the discreteness of the MPS transfer matrix. As we illustrate in the Appendix \ref{sec:app}, such scale separation might not be accessible in the MPS simulations. The above approach becomes viable at the critical point where there is no physical length scale that has to be respected.

\section{Summary of the approach}
\label{sec2}

In this article we employ consecutive largest TM eigenvalues to quantify the divergence from the continuous spectrum necessary to capture the algebraic part of the asymptotics.
In the simplest case we use the distance between the third and second eigenvalue, i.e.,
\begin{equation}
\label{eq:delta0}
\delta = \epsilon_2 - \epsilon_1
\end{equation}
as a refinement parameter that measures the deviation from the exact solution. If needed, the above simple measure can be further refined by taking into account the fact that some form factors may vanish, the transfer matrix can be degenerate and that the system might display some symmetries. The above situations are summarized below and discussed in the subsequent sections of the article. 

We calculate $\epsilon_1(D)$ and $\delta(D)$ for a few MPSs obtained for different bond dimensions $D$, where we observe that the dependence is usually smooth and regular. This allows us to extrapolate $\delta \to 0$ in order to extract the true value of the correlation length with good precision. We compare this approach with the one where $1/D$ is used as a refinement parameter. We observe that $\epsilon_1$ is significantly less regular as a function of $D$ -- especially away from the critical point --  and the result of the extrapolation is less reliable. 
The refinement parameter defined in Eq.~\eqref{eq:delta0} proves to be a good starting point, being sufficient in many simpler cases. However, one of the advantages of quantifying the distance from the exact solution using intrinsic quantities calculated for a given MPS approximation -- in contrast to external parameter such as the bond dimension -- is that we can easily take into account additional information about the state to further refine it if necessary.  This allows to uncover additional information that the MPS description is carrying as well as increase the precision.

Firstly, let us focus only on the part of the transfer matrix spectrum relevant for some particular correlation function $C_{ o  q}(R)$. To that end, we take into account only those transfer matrix eigenvalues for which the corresponding form factors $f^{oq}_j$ are nonzero (within numerical precision), or dominant as compared to the other ones. We mark such eigenvalues with a tilde and additional superscript, $\tilde \epsilon^{oq}_k$. In such cases, we define the refinement parameter as
\begin{equation}
\label{eq:deltaXX}
\delta =\tilde  \epsilon^{oq}_2 - \tilde \epsilon^{oq}_1.
\end{equation}
We apply this definition throughout the article as the most reliable indicator of which TM eigenvalues are relevant and should be taken into account.

Secondly, dominant TM eigenvalues are usually found in groups with well-defined complex phases corresponding to periods of oscillation of the correlation function. The nontrivial correspondence between those phases and the minima of the dispersion relation of the Hamiltonian for which a given MPS is the ground state is discussed in Ref.~\onlinecite{Zauner_2015}.
In order to define the refinement parameter, we can focus only on part of TM spectrum with a given complex phase $\varphi$:
\begin{equation}
\label{eq:deltaphi}
\delta = \tilde \epsilon^\varphi_2 - \tilde \epsilon^\varphi_1.
\end{equation}
We discuss this approach further in Sec.~\ref{sec3a}, where we study the incommensurate phase of the XY model.

Thirdly, there are models for which the dominant TM eigenvalues are either degenerate or are effectively becoming degenerate with the increasing bond dimension. In such a case it is necessary to define the refinement parameter as 
\begin{equation}
\label{eq:deltan}
\delta = \epsilon_n - \epsilon_1,
\end{equation}
where the eigenvalues $\epsilon_1,\ldots,\epsilon_{n-1}$ are (near) degenerate. We observe that behavior e.g. for the XXZ spin-$\frac12$ model in Sec.~\ref{sec3b}, where 4 dominant eigenvalues are near degenerate and we use $n=5$ in the definition above. More generally, even without degeneracy, any definition of the refinement parameter $\delta=\epsilon_n-\epsilon_1$ with $n>1$ should lead to the same extrapolated value of the correlation length. In practice, however, we observe that using smaller $n$ allows for more precise results.

Finally, if the system has some local symmetry and MPS is implemented to take it into account, then the TM spectrum splits into groups with well-defined symmetry charge $u$. We can define the refinement parameter using only 
the eigenvalues belonging to one symmetry sector:
\begin{equation}
\label{eq:deltau}
\delta = \tilde \epsilon^u_2 -\tilde  \epsilon^u_1.
\end{equation}
This is equivalent to selecting a particular subset of correlators corresponding to a given symmetry sector. We use the above approach for the XXZ and Bose-Hubbard models in Secs.~\ref{sec3b}--\ref{sec3d}, which all have $U(1)$ symmetry.

It should be pointed out that the above features are not independent and can be used simultaneously. Ultimately, the reliability of the extrapolation hinges on the consistency of the data obtained for different bond dimensions and through application of different refinement features.

\begin{figure*} [t]
\begin{center}
 \includegraphics[width= 0.68\columnwidth]{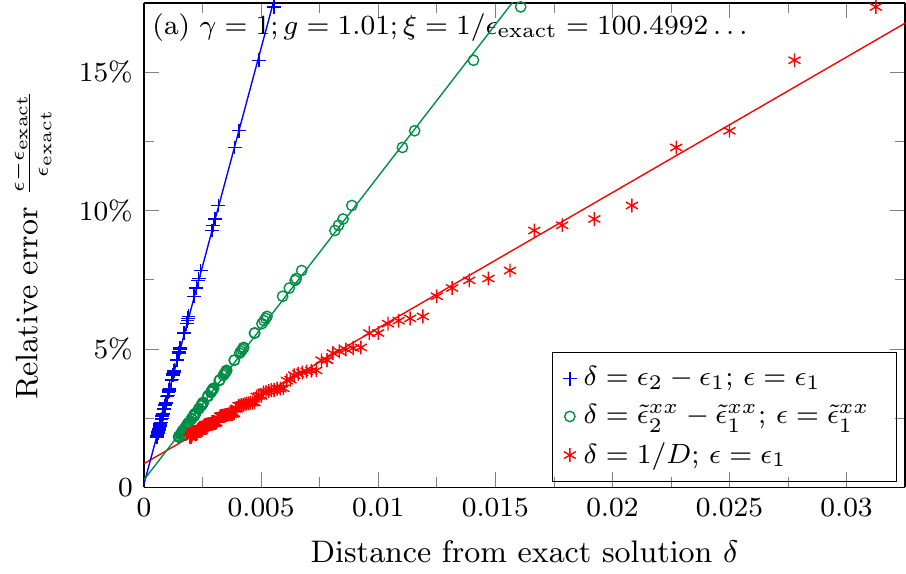}
   \includegraphics[width= 0.68\columnwidth]{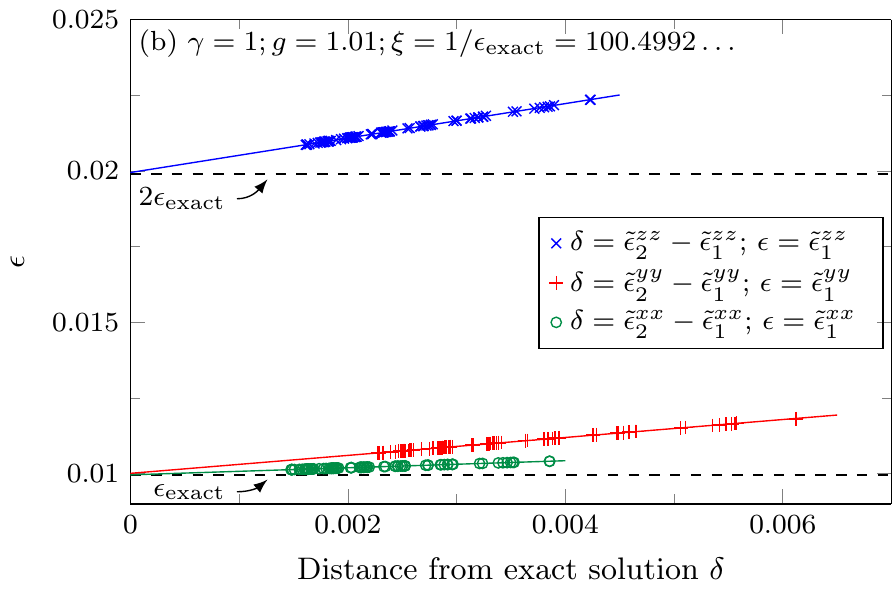}
 \includegraphics[width= 0.68\columnwidth]{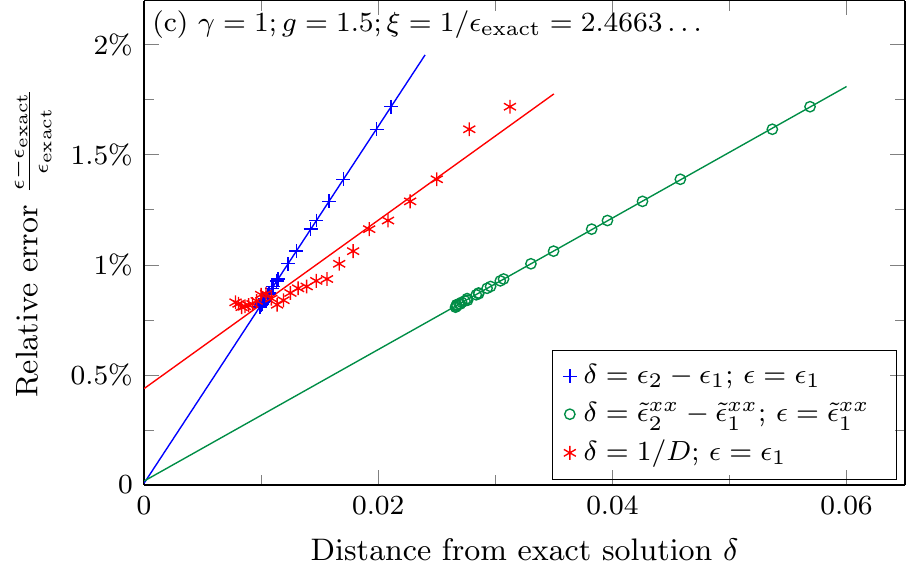}
   \includegraphics[width= 0.68\columnwidth]{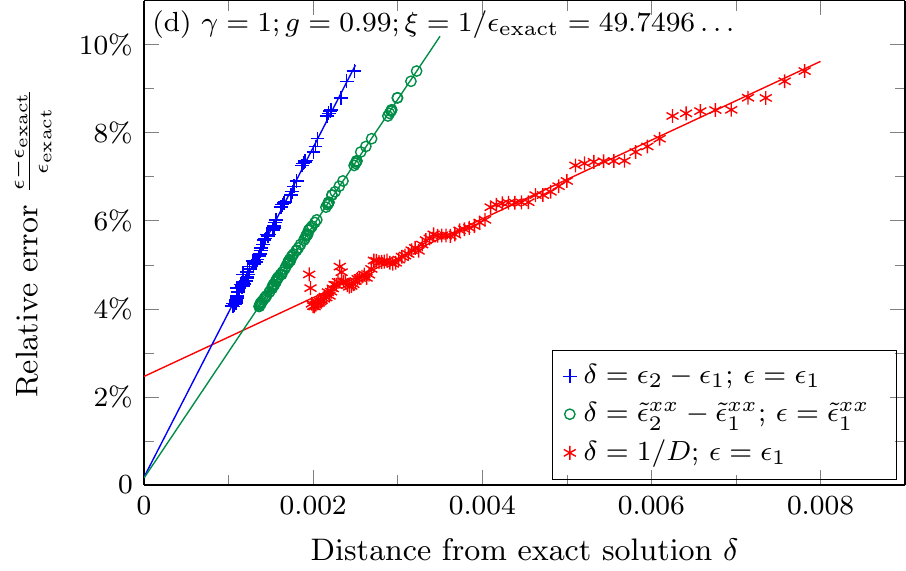}
   \includegraphics[width= 0.68\columnwidth]{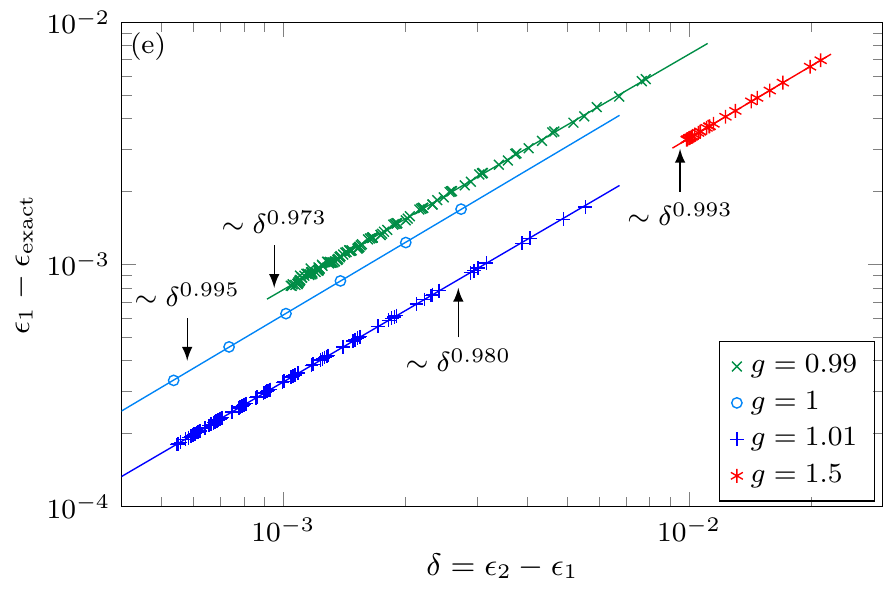}
   \includegraphics[width= 0.68\columnwidth]{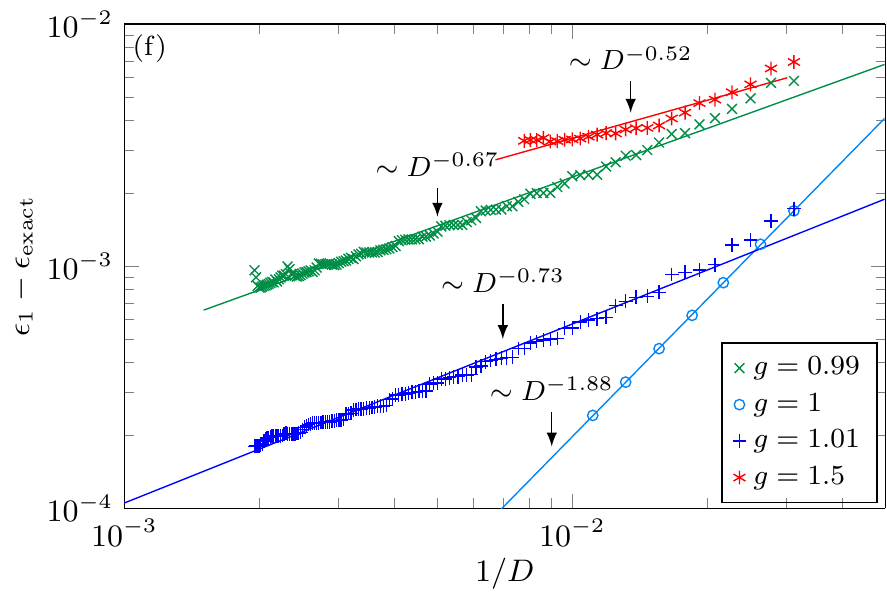}
\end{center}
  \caption{ Extrapolation of the correlation length in the Ising model, $\gamma=1$. See text for discussion. Panel (a): paramagnetic phase with $g=1.01$ and results for bond dimensions $D = 32$---$512$ with step $dD = 4$. Points represent numerical data and solid lines correspond to linear fits.  Panel (b): for $g=1.01$ each TM eigenvalue (or at least a few dominant ones) have nonzero form factor corresponding to exactly one of the correlators $C_{xx}(R)$, $C_{yy}(R)$, $C_{zz}(R)$. This allows to distinguish that the correlation length associated with $C_{zz}(R)$ is halved as compared to the other ones. Here, $D = 128$---$512$. Panel (c): Results for $g=1.5$ show that even far away from the critical point $g_c=1$, extrapolation is necessary to 
 precisely recover the correlation length. Here, $D = 32$---$128$. Panel (d): results for $g=0.99$ in the ferromagnetic phase. Panels (e,f): Log-log plots show error $\epsilon_1 - \epsilon_{\rm exact}$ as a function of refinement parameter $\delta$ suggested in this article, as well as a function of $D$. This validates the general extrapolation model in Eq.~\eqref{eq:extrapolate} and further shows that the  extrapolation method based on $D$  is less useful.}
   \label{fig:TM_Ising}
\end{figure*}

Let us now comment on the extrapolation model that we employ. By analyzing a number of exactly solvable systems we observe that very good results are obtained if one extrapolates by fitting the function
\begin{equation}
\label{eq:extrapolate}
\epsilon =   \epsilon_e +  a \delta^{b},
\end{equation}
where $1/\epsilon_e$ is the extrapolated value of the correlation length and where we assume
 that the error of the inverse of the correlation function, $\epsilon - \epsilon_{\mathrm{exact}}$, is vanishing as a power law with $\delta$.  The exponent $b$ is usually slightly smaller than $1$ and in many cases linear fit, i.e. fixing $b=1$ in Eq.~\eqref{eq:extrapolate} proves to be sufficient. It is also a good starting point, which can then be further tested by allowing $b \neq 1$ and checking if this significantly improves the quality of the fit \footnote{In practice, we perform linear fits for fixed values of $b$, in the end picking the one that minimizes the sum of residuals squared.  Subsequently, we use the values obtained in such a way as initial parameters for nonlinear fit which allows to avoid getting  stuck in local minima. We use the standard nonlinear fitting toolbox as implemented in MATLAB.}.  We use $95\%$ confidence bounds from the non-linear fit in order to estimate an error of extrapolation. We observe that it provides a sensible measure of the quality of the result.


\section{Matrix product states simulations}
\label{sec3}
In this section we benchmark our approach in a range of models of increasing difficulty. As we move to more difficult models, we illustrate different ways of introducing refinement parameters as briefly discussed in the previous section.
 
\subsection{XY model} 
\label{sec3a}

We start with the one-dimensional XY model
\begin{equation}
H = -\sum_m \left( \frac{1+\gamma}2 \sigma^x_m \sigma^x_{m+1} +  \frac{1-\gamma}2 \sigma^y_m \sigma^y_{m+1} + g \sigma^z_m \right),
\label{eq:HXY}
\end{equation}
with anisotropy parameter $\gamma$ and magnetic field $g$. This model is exactly solvable and the asymptotic form of the connected correlation functions is long known \cite{barouch_statistical_1971}. We cite the relevant results below. The numerical results in this section were obtained using the variational uniform matrix product states algorithm (VUMPS) of Ref.~\onlinecite{zauner_variational_2017}, with one-site unit cell, which is based on the time-dependent variational principle approach \cite{haegeman_tdvp_2011, haegeman_unifying_tdvp_2014}. All the states for different bond dimensions were converged with the norm of the energy gradient below $10^{-12}$. Similarly, the maximal change of the Schmidt values in the last iterations of the algorithm (which is another strict measure of convergence) was of the same order.


\subsubsection{Ising model}

We start with the Ising model by setting $\gamma = 1$ in Eq.~\eqref{eq:HXY} and note that the TM spectrum is real and positive in this case. We collect the numerical results in Fig.~\ref{fig:TM_Ising} and discuss them below.

In the paramagnetic phase, for $g>1$, the correlation functions behave asymptotically as 
$ C_{xx}(R) \sim R^{-1/2}  e^{- R / \xi}$,  $ C_{yy}(R) \sim R^{-3/2} e^{- R / \xi}$ and  $C_{zz}(R) \sim R^{-2} e^{-2 R / \xi}$, where the inverse of the correlation length
$1/\xi = \epsilon_{\mathrm{exact}} =  \ln g$. Note the additional factor of two in the exponential part of $C_{zz}(R)$ which is halving the correlation length that appears there.
The results for $g=1.01$ are shown in Fig.~\ref{fig:TM_Ising}(a). Several observations are in order. Without extrapolation, even for the relatively large bond dimension $D=512$ for which the smallest Schmidt value of the MPS bipartition is of the order of $10^{-14}$, the relative error of the correlation length is still $\simeq 2\%$. 

The dependence of  $\epsilon_1$ on $\delta = \epsilon_2-\epsilon_1$, Eq.~\eqref{eq:delta0}, is close to linear. This behavior is already seen for the smallest bond dimensions presented on the plot. Linear regression allows for extrapolation of the true correlation length to within relative error below $0.1 \%$, which can be made even smaller by neglecting the smallest $D$ shown in the picture. It is interesting to note that the eigenvalues $\epsilon_1$ and $\epsilon_2$, used here to calculate the distance $\delta$ above, contribute to different correlators $C_{xx}(R)$ and $C_{yy}(R)$, respectively. Nevertheless, this approach proves to work well in this model. We obtain a consistent result and similar accuracy when we take into account  nonzero form factors, and consider only the part of the TM spectrum that contributes to $C_{xx}(R)$, i.e. using Eq.~\eqref{eq:deltaXX}.

The above can be contrasted with direct application of the bond dimension as a refinement parameter, where the first natural choice is $\delta = 1/D$. In this case  $\epsilon_1$ is oscillating as a function of $1/D$, making extrapolation significantly less reliable as it becomes arbitrary which points to choose for extrapolation. Linear regression for $D=32$---$512$ recovers the correlation length with relative error $\simeq 1\%$, more than an order of magnitude worse than our approach. 
Indeed, by exploring solvability of the model and in particular its Schmidt spectrum,  it was argued in Ref. \onlinecite{Rams_2015} that $\epsilon_1$ should be approaching the exact value much slower than linearly in $1/D$. This explains why linear regression is still underestimating the true value of $\xi$ -- a feature which for sufficiently large $D$ is shared by all the models studied in this article.

In Fig.~\ref{fig:TM_Ising}(b)  we focus on parts of TM spectrum contributing to different correlators: $C_{xx}(R)$, $C_{yy}(R)$ and $C_{zz}(R)$. In the case of a paramagnetic Ising model we observe that each eigenvalue has exactly one form factor which is nonzero. It is either $f^{xx}_j$, $f^{yy}_j$ or $f^{zz}_j$. This allows to recover the fact that the correlation length associated with $C_{zz}(R)$ is halved as compared to the other two, in agreement with the exact result. This shows that such information is encoded, and can be directly extracted from the MPS TM. It is worth noticing that the dominant part of the TM spectrum is relatively sparse: all points in Fig.~\ref{fig:TM_Ising}(b) were obtained using the information from up to 11 largest TM eigenvalues and this was enough to distinguish and extrapolate two correlation lengths differing by a factor of two, even for the largest $D=512$ used there.

In Fig.~\ref{fig:TM_Ising}(c), we show the results for  $g=1.5$, illustrating that the problem with precise extrapolation is present even far away from the critical point when the correlation length is of the order of few sites only. Even in this simple case, without resorting to extrapolation, it is virtually impossible (as the smallest Schmidt values are falling below numerical precision) to recover the true correlation length with relative error below $0.75\%$. On top of that the value of the relative error, say for fixed $D$, clearly depends on the distance from the critical point -- compare with Fig.~\ref{fig:TM_Ising}(a) for $g=1.01$. This makes any fits that use correlation length obtained directly from MPS with fixed $D$, e.g., extracting critical exponents or the position of the critical point out of it, much less trustworthy. Proper extrapolation, as suggested in this article, allows to significantly mitigate this problem.

In Fig.~\ref{fig:TM_Ising}(d), we show results for the Ising model in the ferromagnetic phase with $g=0.99$. In the regime $0<g<1$, the correlation functions behave asymptotically as $C_{xx}(R) \sim R^{-2} \exp(- R/\xi)$,  $C_{yy}(R) \sim R^{-3} \exp(- R/\xi)$, and $C_{zz}(R) \sim R^{-2} \exp(- R/\xi)$ with the inverse of the correlation length $1/\xi = \epsilon_{\mathrm{exact}} =  -2 \ln g$. All the observations made for the paramagnetic phase above fully apply here as well. 

Finally, in Fig.~\ref{fig:TM_Ising}(e) we show the validity of the general extrapolation ansatz introduced in Eq.~\eqref{eq:extrapolate}. The error, that is the distance between $\epsilon_1$ and the exact value, is vanishing as a power law of the refinement parameter proposed in this article with the exponent close to 1. In this simple model, however, we observe that linear regression -- as discussed in the text above -- is sufficient and a more general power law does not result in qualitative improvement of the results in this case. Part of the reason might be that even without extrapolation relative errors are already small here -- at least as compared to other, more complicated models discussed below. More importantly, the exponent $b$ is very close to 1 here. 

When $1/D$ is used as a refinement parameter, the error of $\epsilon$ does not follow a simple functional form, as can be seen in Fig.~\ref{fig:TM_Ising}(f). While it can be locally approximated by a power law, it is evidently flattening, making it a poor ansatz for extrapolation. We note that the critical point is a single exception here, as observed in the context of finite entanglement scaling \cite{Tagliacozzo_FES_2008, Pollmann_FES_2009, Pirvu_FES_2012, Kjall_FES_2013}.


\subsubsection{Incommensurate ferromagnetic phase}
\begin{figure} [t]
\begin{center}
  \includegraphics[width= 0.95\columnwidth]{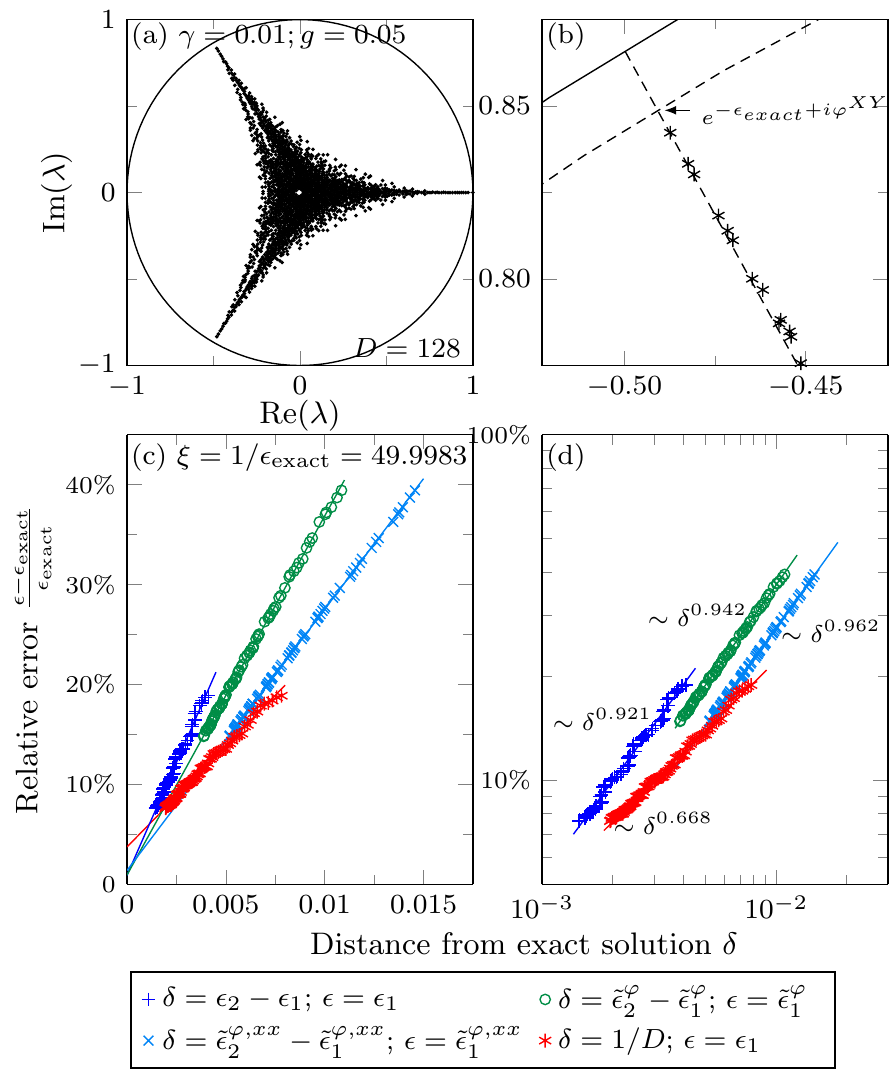}
\end{center}
  \caption{ XY model in the ferromagnetic incommensurate phase, $\gamma=0.01$ and $g=0.5$. Panel (a) shows full TM spectrum for $D=128$. The branch with complex angle $\simeq \varphi^{XY}$ is enlarged in panel (b), where the dashed lines show phase and modulus corresponding to the exact correlation length, respectively.
  We show the results of linear extrapolation in panel (c) focusing on $D=128$---$512$. The same data are plotted in panel (d) in a log-log scale showing the dominant power-law dependence of relative error on the refinement parameter~$\delta$.
  }
   \label{fig:TM_XY_ga001g05}
\end{figure}

In this section we focus on the incommensurate ferromagnetic part of the phase diagram of the XY model, $g^2 + \gamma^2 < 1$, where the correlation function is not vanishing monotonically but has an additional oscillating term. For $g>0$ and $0<\gamma<1$, the leading asymptotics of the  correlation functions is $C_{xx}(R) \sim R^{-2} \exp(- R/\xi)$, $C_{yy}(R) \sim R^{-1} \exp(- R/\xi)$,  and $C_{zz}(R) \sim R^{-2} \exp(- R/\xi)$. In this case the asymptotic behavior may be additionally modified by an oscillating term. Its frequency is given by $\varphi^{XY} =  2 \arccos(g /\sqrt{1-\gamma^2})$.
The correlation length is $1/\xi =\epsilon_\mathrm {exact} = \ln \frac{1+\gamma}{1-\gamma}$. We present the results in Fig.~\ref{fig:TM_XY_ga001g05}.

In Fig.~\ref{fig:TM_XY_ga001g05}(a) we show the full TM spectrum on a complex plane. The dominant eigenvalues form groups with well-defined complex phases \cite{Zauner_2015}, $0$ and $\pm \varphi^{XY}$, respectively. They correspond to frequency of oscillations of the correlation functions. 
 We enlarge the $\varphi^{XY}$ branch in panel (b). It can be seen that the exact phase $\varphi^{XY}$ is well reproduced in the simulation, especially for the dominant eigenvalue. 
In panel (c) we show the results of extrapolation using linear fit. In this model, the simple distance from Eq.~\eqref{eq:delta0} used without any additional refinement results in not-to-smooth functional dependence. Nevertheless, linear regression still reproduces an exact value up to $1\%$. The data are significantly smoother if one focuses only on the part of the TM spectrum with a $\varphi^{XY}$ complex phase, or additionally takes into account nonzero form factors. The error resulting from linear regression is, however, still $\sim 1\%$. All those approaches yield much better results than a linear fit as a function of $1/D$ for which the error is $\sim 4\%$.
This is clarified in panel (d) where we present the data on a log-log plot, showing that the dependence of relative error on $\delta$ is better described by a power law with the exponent that is close to 1 in our approach. 
Indeed, selecting points that have the same complex phase $\varphi^{XY}$ and applying a nonlinear fit allows to reduce the error of extrapolation further to $\sim0.3\%$.


\subsection{XXZ spin-$\frac12$ model}
\label{sec3b}
In this section we analyze the results for spin-$\frac12$ XXZ model
\begin{equation}
H = \sum_m  \left( \sigma^x_m \sigma^x_{m+1} +  \sigma^y_m \sigma^y_{m+1}  + \Delta   \sigma^z_m \sigma^z_{m+1} \right),
\label{eq:XXZ12}
\end{equation}
where we focus on a massive antiferromagnetic regime $\Delta>1$.  The asymptotics of the longitudinal correlation function in this regime was recently calculated  in Ref.~\onlinecite{Dugave_XXZmassive_2015} as $C_{zz}(R) \sim R^{-2} e^{-R/\xi}$, see Eq.~(3.36) therein for the full expression.  The algebraic part of the transverse correlation function $C_{xx}(R)$ is characterized by the same exponent \footnote{J.~Suzuki (private communication). The asymptotics of $C_{xx}(R)$ can be derived using the results presented in Ref.~\cite{Dugave_XXZmassive_2016} \label{Suzuki_private}}.
The correlation length in the model reads $1/\xi = \epsilon_{\mathrm{exact}} = -\log(k(q^2))$, where $q = e^{-\mathrm{arccosh} \Delta}$ and the elliptic modulus $k(q^2) = \vartheta^2_2(0,q^2)/\vartheta^2_3(0,q^2)$ with $\vartheta_n(z,q)$ being Jacobi theta function \cite{Dugave_XXZmassive_2015, Johnson_8vertex_1973, Dugave_XXZmassive_2016}.

\begin{figure} [t]
\begin{center}
  \includegraphics[width= 0.95\columnwidth]{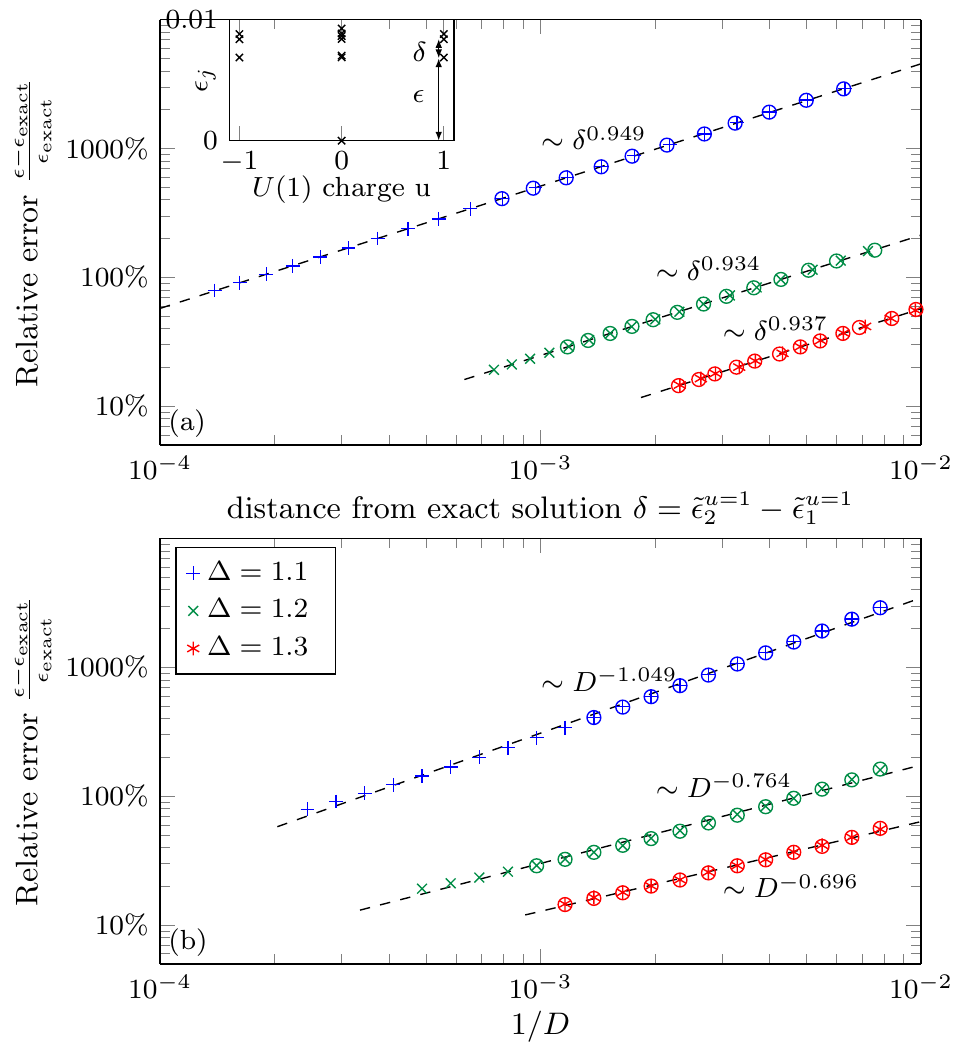}
\end{center}
  \caption{ XXZ spin-$\frac12$ model. Panel (a): Stars correspond to the error as a function of refinement parameter $\delta = \tilde \epsilon^{u=1}_2-\tilde \epsilon^{u=1}_1$, where we employ $U(1)$ symmetry and $u$ is the symmetry charge. Inset shows splitting of  the dominant part of TM spectrum into symmetry sectors for   $\Delta=1.2$ and $D=2048$. Circles show the data obtained without employing symmetries and $\delta = \epsilon_5-\epsilon_1$; note near degeneracy of the  dominant four TM eigenvalues in the inset. Panel (b): The same data with the modification that $\delta=1/D$ is used as a refinement parameter. Only  points for $D\ge128$ are shown.}
   \label{fig:TM_XXZ12}
\end{figure}

The numerical results in this and in the following sections were obtained using the iDMRG algorithm \cite{McCulloch_idmrg_2008} with a two-site unit cell incorporating U(1) symmetry \cite{Singh_sym_2010, Singh_U(1)_2011}, which in this case corresponds to the conservation of $S^z_{\rm total}=\sum_m \sigma^z_m$.  All points were converged up to maximal change of the Schmidt values in the last iteration below $10^{-10}$. The bond dimensions $D$ approximately form a geometric series with a step $2^{1/4}$.

We collect the results in Fig.~\ref{fig:TM_XXZ12}, where we focus on the part of the TM spectrum corresponding to $U(1)$ charge $u=1$ with the refinement parameter $\delta$ defined in this sector according to Eq.~\eqref{eq:deltau}. We present the splitting of the dominant part of the TM spectrum for a single $D$ in the inset of panel (a). Figure \ref{fig:TM_XXZ12} shows the data in a log-log scale to highlight power-law dependence of the error of $\epsilon$ on $\delta$. 
For completeness, in panel (b) we show the dependence of the error on $1/D$.  While it is smooth, it does not seem to follow a clear functional form again making it not very useful for precise extrapolation.

We also run simulations with VUMPS with o one-site unit cell \footnote{We used the VUMPS algorithm with the one-site unit cell  and simulate the Hamiltonian equivalent to Eq.~\eqref{eq:XXZ12} but with every second spin rotated to make the dominant interaction in the $z$ direction ferromagnetic. Otherwise, the model spontaneously breaks translational symmetry making the one-site unit cell ill suited to handle such a case.} and without U(1) symmetry where we used the refinement parameter $\delta = \epsilon_5 - \epsilon_1$, as defined in Eq.~\eqref{eq:deltan}, to take into account near degeneracy of $\epsilon_1,\ldots,\epsilon_4$ -- see inset of Fig.~\ref{fig:TM_XXZ12}(a). We present those results to show how to deal with degeneracies when they become an issue for the simplest refinement parameter in Eq. (\ref{eq:delta0}). Those results are represented by circles in Fig.~\ref{fig:TM_XXZ12}. An alternative approach would be to take the suitable form factors into account. Notice that $f^{zz}$ can be nonzero only for eigenvalues belonging to $U(1)$ charge $u=0$. Similarly, $f^{xx}= f^{yy}$ can be nonzero only when $u=\pm 1$.  

Finally, we collect results of the actual nonlinear fits  in Table \ref{tab:XXZ12}. The values are averaged over range of bond dimensions taken into account, where we use 8 to 13 points with largest $D$. Additionally, we also take into account dominant form factors. This is especially relevant for the $u=0$ sector where there is near degeneracy of two dominant eigenvalues, which still has to be resolved (see inset of Fig.~\ref{fig:TM_XXZ12}(a)). Apart from $\Delta=1.1$, the exact value of the correlation length is recovered with the error well below $1\%$.  It is $\sim 3\%$ for $\Delta=1.1$, however the correlation length is approaching $10^4$ here and we extrapolate from MPS correlation lengths underestimating the exact one by almost a factor of 2. Obtaining such results from the simulation of a finite system is practically impossible, showing the effectiveness of the infinite uniform approach.

\begin{table}[t]
\begin{tabular}{|l| l| l| l| l| l|}
 \hline 
  $\Delta$ & $1.1$ & $1.2$ & $1.3$ & $1.4$ & $1.5$\\
  \hline
   $\xi_{\mathrm{exact}}$ & $8482.801$ & $347.131$ & $85.1433$ & $37.0497$ &  $21.0729$\\
   \hline
   $\xi^{xx}_{u=1}$ & $8280(130)$   &  $345.9(24)$   &  $85.54 (99)$ &  $  37.0 (10)$ &   $ 20.89(74) $ \\
  \hline
   $\xi^{zz}_{u=0}$ & $ 8200(120)$   &  $ 348.6(21)$   &  $85.5(12)$ &  $   37.36 (80)$ &   $ 21.12(51) $ \\
  \hline
   $D_{\mathrm{max}}$ & $  4096$   &  $ 2048$   &  $862$ &  $  430$ &   $ 256 $ \\
  \hline
   $\xi_{D_{\mathrm{max}}}$ & $ 4734 $   &  $ 291.3$   &  $74.29$ &  $  32.49$ &   $ 18.51 $ \\
  \hline
\end{tabular}

\caption{XXZ spin-$\frac12$ model. Comparison of the extrapolated correlation length with the analytical result.  Correlation length is resolved by symmetry sector, where $u=1$ can be associated with $C_{xx}(R)$ and  $u=0$ with $C_{zz}(R)$. Additionally, we show the largest bond dimension $D$ used for given $\Delta$ and the corresponding MPS correlation length. For those values of $D$ the exact ground state energy is reproduced up to an error of $O(10^{-14})$ for $\Delta \ge 1.2$ and $O(10^{-12})$ for $\Delta = 1.1$}
\label{tab:XXZ12}
\end{table}

We finish this section with an observation that the TM spectra which we obtain from numerics in this model are real.  There are both positive and negative eigenvalues for the one-site unit cell implementation (complex phases $\varphi=0,\pi$) which corresponds to the monotonic and staggered part of the correlation function asymptotics -- see  Eq.~(3.36) in Ref.~\onlinecite{Dugave_XXZmassive_2015}. This information about the phase is lost in the two-site unit cell transfer matrix, where the spectrum is effectively squared and strictly positive, which shows some advantage of using as small a unit cell as possible.  It is interesting to contrast this with the spectrum of the quantum transfer matrix, which is complex with the complex phase changing in a continuous way; see e.g., Ref.~\onlinecite{Johnson_8vertex_1973} for an in-depth discussion. In that case the minimal gap of the quantum transfer matrix is not dictating the actual correlation length as contributions of part of the QTM spectrum band are effectively canceling out. Apparently, MPS is capturing only the physically relevant part of the spectrum making it purely real in our case. A similar situation -- though slightly more complicated --  arises in the 8-vertex model, which we discuss below in the context of the CTM algorithm.


\subsection{XXZ spin-$\frac32$ model}
\label{sec3c}
In this section, we consider spin-$\frac32$ XXZ model 
\begin{equation}
\label{eq:XXZ32}
H = \sum_m  \left(   S^x_m  S^x_{m+1} +  S^y_m  S^y_{m+1}  + \Delta   S^z_m S^z_{m+1} \right),
\end{equation}
where $ S^{x,y,z}_m$ are standard spin-$\frac32$ operators acting on site $m$.  Similarly to the spin-$\frac12$ case discussed in the previous section, the model has a critical region for $-1 \le \Delta \le 1$ with a BKT critical point at $\Delta_c = 1$  separating the gapped phase for $\Delta > 1$ \cite{XXZ32_1986_Schulz, Affleck_1987, Affleck_1989, XXZ32_Hallberg_1996}. Again, we focus on the latter, where the correlation length scales as
\begin{equation}
\xi(\Delta) = \xi_0 \exp( B / \sqrt{|\Delta-\Delta_c|}).
\label{eq:xi_BKT}
\end{equation}

\begin{figure} [t]
\begin{center}
  \includegraphics[width= 0.95\columnwidth]{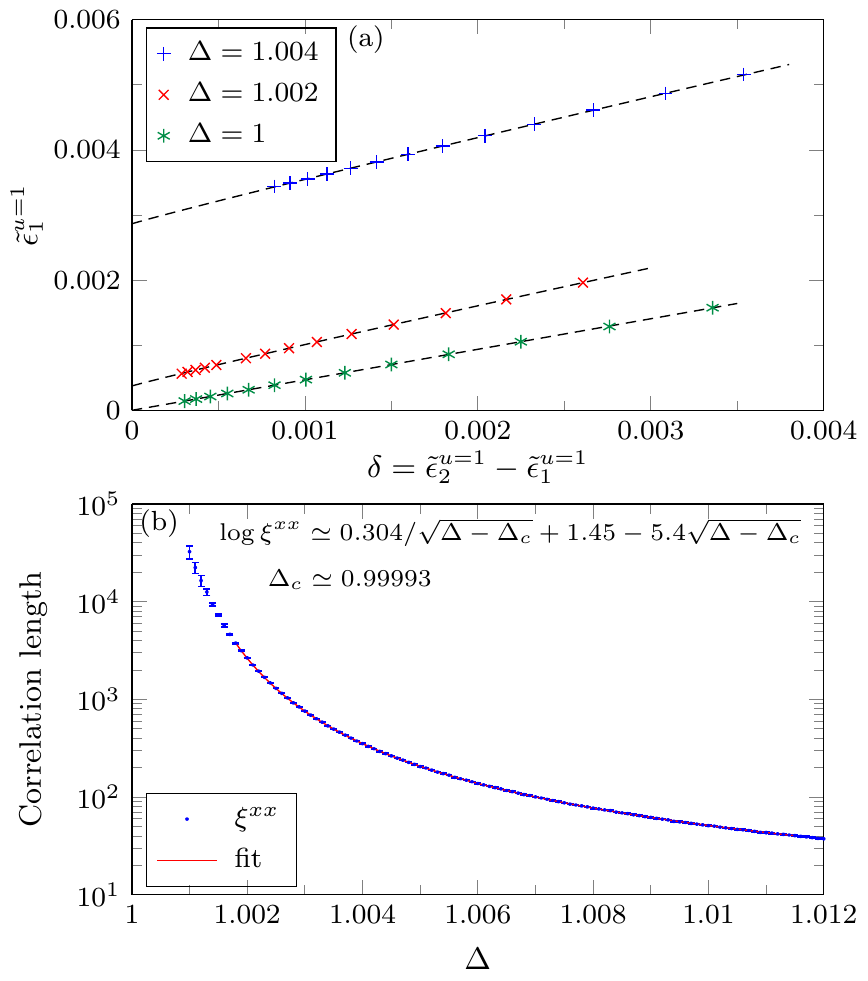}
\end{center}
  \caption{ XXZ spin-$\frac32$ model. Panel (a) shows extrapolation of the correlation length for selected values of $\Delta$. Refinement parameter $\delta$ is calculated from sector with U(1) charge $u=1$, Eq.~\eqref{eq:deltau}. In panel (b) we collect correlation lengths obtained from our extrapolation procedure. Subsequently, we fit the form of $\log(\xi(\Delta))$ given in Eq. (\ref{eq:xi_BKT}). This allows us to recover the exact position of the critical point, $\Delta_c=1$, with excellent accuracy. See text for details.} 
\label{fig:TM_XXZ32}
\end{figure}

Even though the exact value of the critical point is known, the model cannot be solved exactly and the value of the correlation length is not known analytically. Additionally, it is particularly challenging to approximate the ground state due to very strong quantum fluctuations \cite{XXZ32_Rigol_2015, XXZ32_Taddia_2012}. As such, the model  provides a good test for numerical methods. We extrapolate the correlation lengths using our method and subsequently fit the scaling form of Eq.~\eqref{eq:xi_BKT} (with higher-order corrections) in order to extract $\Delta_c$ and the parameter $B$. We collect the results in Fig.~\ref{fig:TM_XXZ32} and a few selected correlation lengths in Table~\ref{tab:XXZ32}. 

We proceed similarly as in the previous section. We define the refinement parameter by  taking into account information about the symmetry sector, Eq.~\eqref{eq:deltau}, and extrapolate by fitting the general power law in Eq.~\eqref{eq:extrapolate} to $8$---$13$ points with the largest bond dimension $D$ used in the simulation. Consecutive $D$ form a geometric series with a step $2^{1/4}$ approaching $10000$ for the most challenging points. We show some of such fits in Fig.~\ref{fig:TM_XXZ32}(a). This also includes the critical point at $\Delta=1$, where we obtain the extrapolated value of the inverse correlation length as $\epsilon_e \sim 10^{-6} \pm 7\cdot10^{-6} $, which is 2 orders of magnitude smaller then the value of $\epsilon_1$ obtained for the largest $D=5792$ converged here. This shows both the limits and validity of our approach, as within the estimation error we are able to recover the exact value (zero) at the critical point.

Subsequently, in  Fig.~\ref{fig:TM_XXZ32}(b), we fit
$\ln \xi = a_1 + B (\Delta-\Delta_c)^{-1/2}+a_2 (\Delta-\Delta_c)^{1/2}$ where we allow for subleading correction with nonzero $a_2$ to improve the quality of the results.  We take into account $\xi \in [40,4000]$, where the estimated errors of extrapolations are below $1.5 \%$.
We obtain $\Delta_c=0.99993(4)$, which is in very good agreement with the exact value of $\Delta_c=1$. The nonuniversal constant $B=0.304(12)$, $a_1 = 1.45(20)$, and $a_2 = -5.4(12)$.

We compare those results with the ones recently reported in Ref.~\onlinecite{XXZ32_Rigol_2015}, which were obtained by fitting the scaling ansatz capturing behavior of the energy gap in the finite system. Namely, $\Delta_c = 0.995\pm0.004$ and $B=0.50\pm0.02$ ($\Delta_c = 0.989\pm0.01$ and $B=0.58\pm0.04$) for open (periodic) boundary conditions, where the systems up to 280 (72) spins were used.
We are able to access the range of correlation lengths which are an order of magnitude larger then system sizes possible in state-of-the-art finite system simulations and also take into account subleading correction to the scaling -- which become increasingly important at a distance from the critical point. As such, we expect our results to be more accurate, which can be seen in the precision with which we were able to localize the critical point.

\begin{table}[t]
\begin{tabular}{|l| l| l| l| l|  l| }
 \hline 
  $\Delta$  & $1.002$ & $1.004$ & $1.006$ & $1.008$& $1.01$\\
   \hline
   $\xi^{xx}_{u=1}$  & $ 2654(26) $ & $351.7(27) $ & $ 137.59(94) $ & $ 76.89(92)$ & $50.90(63) $\\
  \hline
   $D_{\mathrm{max}}$  & $ 9742 $ & $8192 $ & $ 3444   $ & $  2896$ & $512$\\
  \hline
   $\xi_{D_{\mathrm{max}}}$ & $ 1773 $ & $290.9 $ & $ 115.8 $ & $   65.9 $ & $41.3$\\
  \hline

\end{tabular}
\caption{Extrapolated correlation lengths for the XXZ spin-$\frac32$  model. Additionally, we show maximal bond dimension $D$ used in the simulation, as well as MPS correlation length for that $D$.}
\label{tab:XXZ32}
\end{table}


\makeatletter
\setlength{\@fptop}{5pt}
\makeatother

\begin{table*}[th]
\begin{tabular}{| l| l| l| l| l| l|  l| l| l| l| l|}
 \hline 
  $J$ & $0.28$ & $0.27$ & $0.26$ & $0.25$ & $0.24$& $0.23$&$0.22$&$0.21$&$0.20$\\
   \hline
   $\xi^{b^\dagger b}_{u=1}$ & $4235(17)$ & $	772.8(32) $ & $252.6(10) $ & $112.6(10)$ & $ 60.23(29)$ & $36.39(16)$&$23.93(11)$&$16.69(10)$&$12.19(11) $\\
  \hline
  $\xi^{nn}_{u=0}$ & $2096(24)$ & $387.4(65) $ & $127.5(25) $ & $56.0(29) $ & $ 30.1(17)$ & $17.92(70)$&$11.99(66)$&$8.37(70) $&$6.01(55)$\\
  \hline
  $D_{\mathrm{max}}$& $4096$ & $2048$ & $1024$ & $512$ & $362$ & $256$ & $256$ & $182$ & $128$ \\
  \hline
  $\xi_{D_\mathrm{max}}$& $3387$ & $677$ & $227$ & $101$ & $54.9$ & $33.4$ & $22.3$ & $15.5$ & $11.4$ \\
  \hline
\end{tabular}
\caption{Extrapolated correlation lengths in the Bose-Hubbard model with unit filling, $\langle  n\rangle =1$. We show the correlation lengths for $\langle  b^\dagger_0  b_R \rangle$ and $\langle  n_0  n_R \rangle$. Notice that the second one is approximately halved as compared to the first one. Additionally, we show maximal bond dimension $D$ used in the simulation and MPS correlation length for that $D$.}
\label{tab:BH}
\end{table*}

\begin{figure} [b]
\begin{center}
  \includegraphics[width= 0.95\columnwidth]{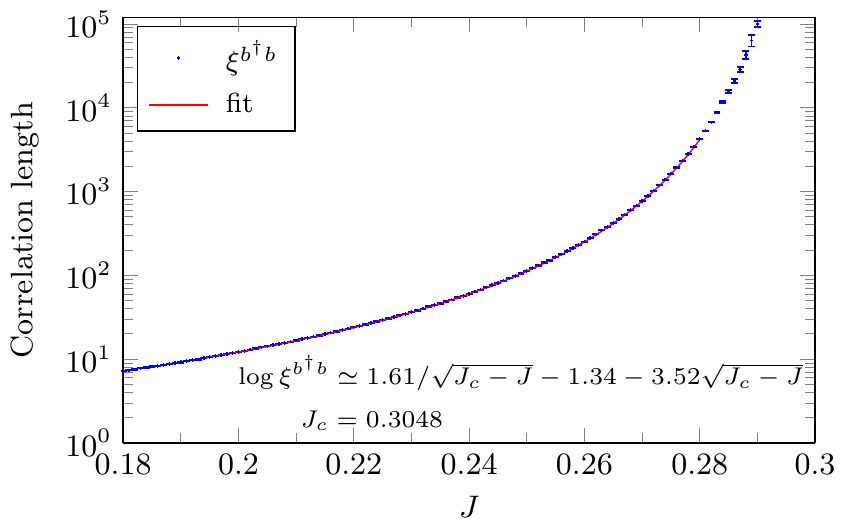}
\end{center}
  \caption{ One-dimensional Bose-Hubbard model with $\langle  n\rangle =1$. We show extracted values of the correlation lengths associated with $\langle b^\dagger_0 b_R \rangle$ correlator. By fitting the scaling form in Eq.~\eqref{eq:xi_BKT}, we find the position of the critical point and nonuniversal constant $B$ in this model. See text for details.} 
\label{fig:TM_BH}
\end{figure}

\subsection{Bose-Hubbard model}
\label{sec3d}

We conclude this part with the Bose-Hubbard model in one dimension,
\begin{equation}
H = - J \sum_m  \left (   b_{m+1}^\dagger   b_m +   b_{m}^\dagger   b_{m+1} \right) + \frac{U}{2} \sum_m     n_{m} (  n_{m} - 1) ,
\end{equation}
where $  b_m$ are bosonic annihilation operators acting on site $m$ and $ n_{m} =  b_m^\dagger   b_m$ is  the particle number operator. Below we set the energy scale by fixing the Coulomb repulsion $U=1$ and consider a system with unit filling per lattice site, $\langle  n_{m} \rangle = 1$.  The model has a quantum phase transition between the gapped Mott insulator phase for $J<J_c$, and the gapless superfluid phase for $J>J_c$ in the  Berezinskii–Kosterlitz–Thouless universality class \cite{Fisher_BH_1989, Krutitsky_BH_review_2016}. We focus on the gapped phase and proceed identically as in the previous sections.

In the iDMRG simulations we truncate the local Fock space at six particles,  checking that this is enough to obtain converged results.  We employ U(1) symmetry, which in this case corresponds to conservation of the total particle number, $\sum_m  n_m$. This model proves to be less challenging for MPS simulations than the XXZ spin-$\frac32$ model from the previous section and all the results were obtained with bond dimension up to $5792$.

We collect the results of our extrapolation procedure in Fig.~\ref{fig:TM_BH} and in Table~\ref{tab:BH}. We calculate both the correlation length associated with $\langle b_0^\dagger  b_R \rangle$ and $\langle n_0  n_R \rangle$ correlators. We observe that $\xi^{nn}$ is halved as compared to $\xi^{b^\dagger b}$ within the estimated extrapolation errors; see Table~\ref{tab:BH}. Subsequently, we focus on $\xi^{b^\dagger b}$, which can be extrapolated with higher accuracy, and fit $\ln \xi^{b^\dagger b} = a_1 + B (J_c-J)^{-1/2}+a_2 (J_c-J)^{1/2}$. We  use $\xi \in [10,5000]$, for which the estimated extrapolation errors are well below $1 \%$. We obtain the position of the critical point as $J_c = 0.3048(3)$ and nonuniversal constant $B=1.61(4)$. Additionally, $a_1 = -1.34 (15)$ and $a_2 = -3.52(24)$. 

For the collection of results on the critical point position obtained in a multitude of different studies, see Table 1 in the recent review article Ref.~\onlinecite{Krutitsky_BH_review_2016}, with the current consensus of $J_c \approx 0.3$. The value obtained in Ref. \onlinecite{Rigol_BH_2013} from studies of the energy gap in the finite DMRG simulation of up to $700$ sites, where the value of $B$ was also reported, reads $J_c=0.3050\pm0.0001$ and $B = 1.59\pm0.03$,  which are in very good agreement with our results.

To conclude, we comment that the range of estimates of the critical point position, as collected in Ref.~\onlinecite{Krutitsky_BH_review_2016}, illustrates very well the complexity of such studies. It emphasizes the necessity of using methods that are unbiased and able to precisely capture extreme values of the correlation length. The former is provided by MPS-based schemes, evidently seen by the consistency of the results obtained using those methods, see Ref.~\onlinecite{Krutitsky_BH_review_2016}. The latter can be provided by working directly in the thermodynamic limit, which allows to avoid problems posed by strong finite-size effects and limitations on the possible system sizes in such simulations. On the other hand, a proper extrapolation scheme of the correlation length significantly lessens the systematic limitations caused by finite bond dimension always present in MPS simulations. Both allow us to expect excellent accuracy of the results presented here, especially as we keep in mind the quality of the data obtained for the XXZ model in the previous section.


\section{Corner transfer matrix simulations}
\label{sec4}
\begin{figure} [t!]
\begin{center}
  \includegraphics[width= 0.95\columnwidth]{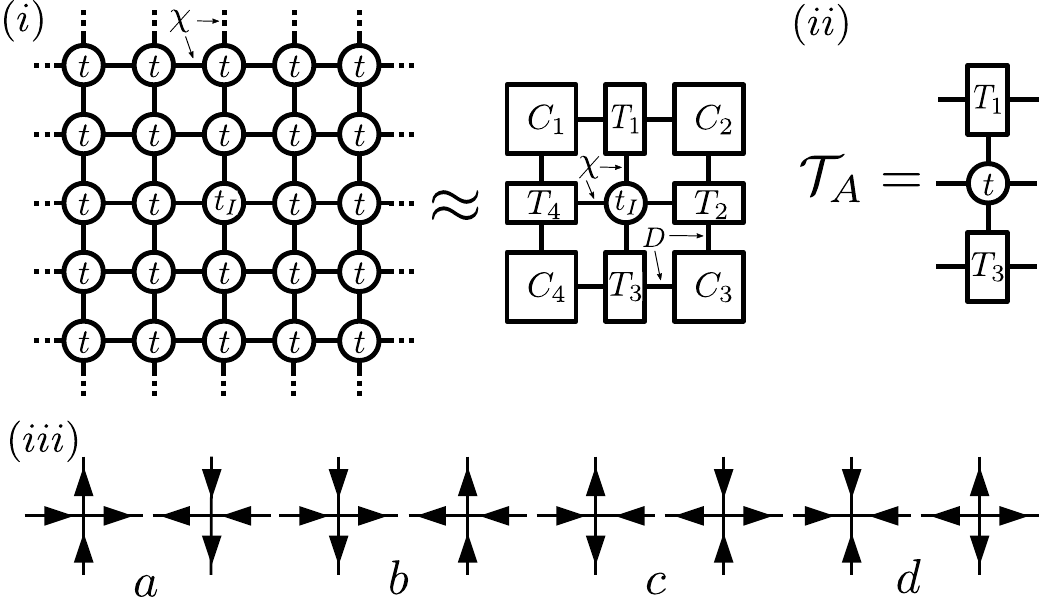}
\end{center}
  \caption{$(i)$ Visualization of the corner transfer matrix renormalization method where the environment of a single site, marked as $t_I$ here, of infinite 2d lattice is approximated by a combination of finite size (finite D) tensors $T$ and $C$.  $(ii)$ Definition of column-to-column transfer matrix used in this work. In $(iii)$ we show allowed configurations and their  Boltzmann weights in the 8-vertex model. They are combined into one tensor $t$ in $(i,ii)$, with $\chi=2$ being a number of possible states of local variable living on the edges of the lattice.} 
\label{fig:def_8vertex}
\end{figure}

In the corner transfer matrix methods the infinite environment of a given site (or sites in a unit cell) in a 2d tensor network is approximated by a combination of four corner $C_j$ and four top $T_j$ tensors of finite size, as depicted in Fig.~\ref{fig:def_8vertex}$(i)$. This allows to compute expectation values of local operators of interest as well as their correlation functions. To that end, we define column-to-column transfer matrix as shown in Fig.~\ref{fig:def_8vertex}$(ii)$.

We employ the general corner transfer matrix algorithm described in Ref.~\onlinecite{Corboz_CTM_2014}, suitable for nonsymmetric problems (i.e., when corners $C_j$ are not Hermitian) and its less expensive variant described in Ref.~\onlinecite{Czarnik_2016}. As compared to the former one, it effectively avoids squaring small Schmidt values of the enlarged corners, which are later inverted in the algorithm. Furthermore, it reduces the leading cost of the algorithm. This allows to reach larger CTM bond dimensions, which we call $D$.


\subsection{Classical 2d Ising model}
The 2d classical Ising model can be exactly mapped on the 1d XY model \cite{Suzuki_1971,Suzuki_1976}, which we discussed in details in Sec.~\ref{sec3a} and as such we are not going to repeat those results here.

The main point that we would like to make here is that the proposed extrapolation procedure gives essentially the same result for two algorithms: the CTM method for 2d classical model and the VUMPS algorithm for the corresponding XY model. The CTM method was used in its symmetric \cite{Nishino_CTM_1996, Nishino_CTM_1997} as well as nonsymmetric \cite{Corboz_CTM_2014, Czarnik_2016} form. We conclude that the accuracy of the extrapolation method described in this article, and more generally renormalization of the exact quantum transfer matrix in CTM/MPS simulations \cite{Zauner_2015, Rams_2015, Bal_2016}, is mostly independent of the particular algorithm used to obtain it.


\subsection{8-vertex model}
In this section we test our approach in a numerically significantly more challenging 8-vertex model. We use the standard formulation of the model, see, e.g. Refs. \onlinecite{Baxter, Johnson_8vertex_1973, Krcmar_8vertex_2016}, with local two-state variables living on the edges of a 2d square lattice.
Each local variable is represented by an arrow pointing toward one of the two adjacent lattice vertices. Only configurations with an even number of arrows pointing out of any vertex are allowed and contribution of each configuration to the partition function
is calculated as a product of Boltzmann weights for each vertex, assigned as in Fig.~\ref{fig:def_8vertex}$(iii)$. The weights are parametrized as
\begin{eqnarray*}
a &=&  e^{(J+J'+J'')/T} \\
b &=&  e^{(-J-J'+J'')/T} \\
c &=&  e^{(-J+J'-J'')/T} \\ 
d &=&  e^{(J-J'-J'')/T}.
\end{eqnarray*}

\begin{figure} [t]
\begin{center}
  \includegraphics[width= 0.95\columnwidth]{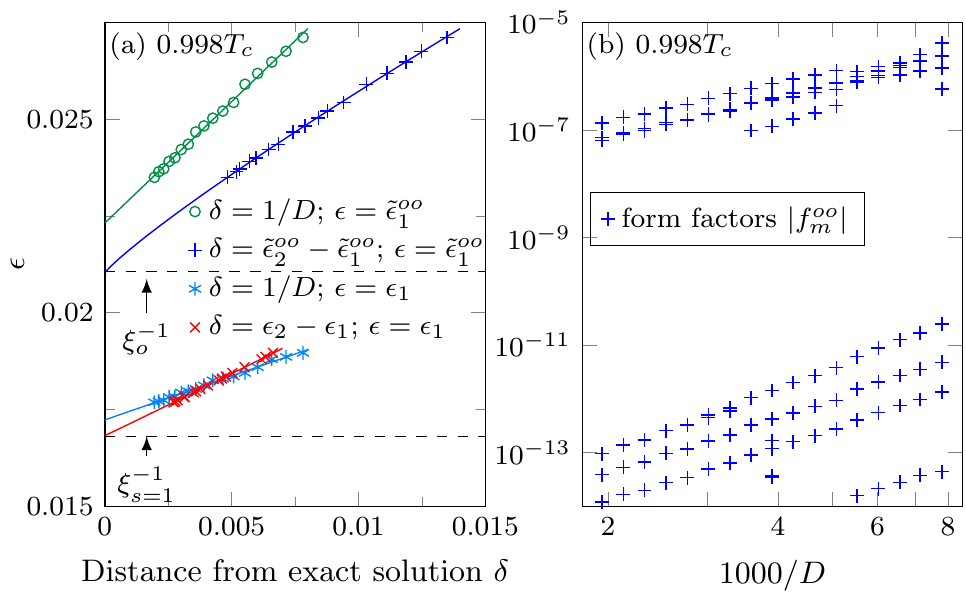}
  \includegraphics[width= 0.95\columnwidth]{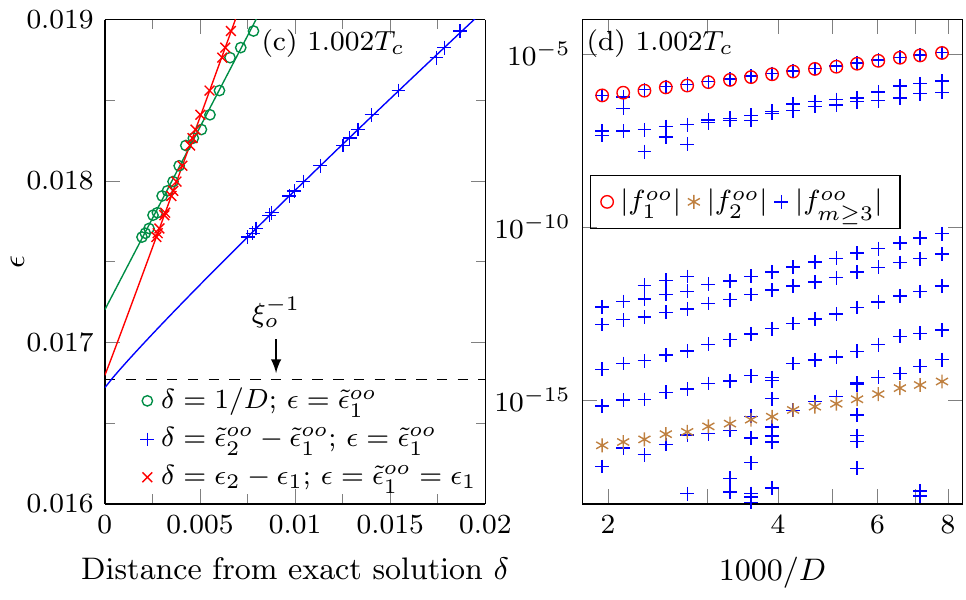}
\end{center}
  \caption{ 8-vertex model with $J=0.2$, $J'=J''=0.1$. Panels (a) and (c) show results of extrapolation for ferroelectric $T=0.998 T_c$ and disordered $T=1.002 T_c$ phases, respectively. Dashed lines indicate analytical results. Panels (b) and (d) show order parameter form factors for the 20 largest TM eigenvalues, $D = 128$---$512$. We use $\tilde \epsilon_n^{oo}$ to mark the TM eigenvalues for which the order parameter form factors are nonzero in order to avoid confusion with $f^{oo}_n$ which is the form factor corresponding to $\epsilon_n$.}
\label{fig:TM_8vertex}
\end{figure}

We perform numerical tests for $J=0.2$, $J'=J''=0.1$ ($a>b,c,d$).
We focus on the vicinity of the continuous critical point, where the critical temperature $T_c$ is set by the relation $a_c = b_c + c_c + d_c$. The system is in the ferroelectric phase for $T<T_c$ and the
disordered phase for $T>T_c$. The expectation value of the vertical arrow direction serves as the order parameter in this model and the correlation length associated with this observable was calculated analytically in Ref.~\onlinecite{Johnson_8vertex_1973}.

As this problem is in general not symmetric (for $c \neq d$), simple CTM implementation \cite{Nishino_CTM_1996} cannot be used \cite{Krcmar_8vertex_2016} and it is  necessary to employ the most general CTM suitable for such problems \cite{Corboz_CTM_2014, Czarnik_2016}. See Ref.~\onlinecite{Krcmar_8vertex_2016} for some recent CTM studies of symmetric, but not exactly solvable, modification of this model. We show the results of simulations in Fig.~\ref{fig:TM_8vertex}, where we have chosen $T=0.998 T_c$ ($\xi_o = 47.4556\ldots$) and $T=1.002 T_c$ ($\xi_o = 59.6230\ldots$), where the correlation length is associated with the vertical arrow correlation along a row.

In the ferroelectric phase, $T=0.998 T_c$, it is necessary to focus on the part of the TM spectrum that contributes to the order parameter correlation function. We define $\delta$ as in Eq.~\eqref{eq:deltaXX}. The correlation length is extrapolated up to an error below $0.1\%$ using nonlinear fit, as shown in Fig.~\ref{fig:TM_8vertex}(a). To that end we used TM eigenvalues for which the form factors plotted in Fig.~\ref{fig:TM_8vertex}(b) are in the $\sim 10^{-7}$ band. The corresponding error when $\delta=1/D$ is used is $\sim 5\%$.

It is worth noting that the TM spectrum contains another, longer scale of length obtained from the two largest TM eigenvalues $\epsilon_{1,2}$. The order parameter form factors corresponding to those eigenvalues  are zero up to the numerical precision, i.e. they are not visible in Fig.~\ref{fig:TM_8vertex}(b) (Note that form factors are defined as a product of two numbers of similar magnitude. This means that the values larger than $10^{-20}$ are considered to be nonzero). 
This length scale corresponds to another band of QTM eigenvalues, which, however, does not contribute to order parameter correlation function as the corresponding form factors vanish due to symmetries of the model \footnote{See Sec. VI.B in Ref.~\onlinecite{Johnson_8vertex_1973}. Our example corresponds to the parameter $\mu \in (\pi/2,2\pi/3]$ used there. The longer length scale, $\xi_{s=1}$, comes from $s=1$ bound states.}.  
In order to break this symmetry we regard the 8-vertex model as a special case of a more general 16-vertex model, where vertices with three-in-one-out and three-out-one-in arrow configurations are considered together with the ones already depicted in Fig.~\ref{fig:def_8vertex}$(iii)$; see e.g. Ref.~\onlinecite{Wu_16vertex_1969, Assis_16vertex_2017, Cugliandolo_16vertex_2017}. $\epsilon_{1,2}$ contributes to the correlation functions, and have nonzero form factors, for operators build from those additional vertices. The corresponding correlation length is represented by the lower dashed line in Fig.~\ref{fig:TM_8vertex}(a). 

In the disordered phase, $T=1.002 T_c$, the situation is much simpler as the dominant eigenvalues have nonzero order parameter form factors. In Fig.~\ref{fig:TM_8vertex}(c) we recover
the true correlation length with error $\sim 0.1\%$, both by focusing on TM eigenvalues with form factors in the dominant band, $\sim10^{-7}$ in Fig.~\ref{fig:TM_8vertex}(d), and by taking $\delta = \epsilon_2 - \epsilon_1$. The second one works well even though the form factor for $\epsilon_2$ (marked as stars in Fig.~\ref{fig:TM_8vertex}(d)) is orders of magnitude smaller than for $\epsilon_1$ (circles).

It is worth observing that the QTM has complex spectrum, and the above physical correlation lengths are shorter than could have been expected from considering only the absolute value of the largest QTM eigenvalues \cite{Johnson_8vertex_1973}. The longest scales effectively cancel out due to a combination of continuously changing complex phase of QTM eigenvalues, together with the  proper symmetry of the form factors (periodicity in the space of parameters quantifying the spectrum). 
On the other hand the spectrum of the column-to-column TM is real and positive. It directly describes the physical length scales, which (if necessary) can be distinguished with the help of nonzero form factors corresponding to proper operators.

  \begin{figure} [b]
\begin{center}
  \includegraphics[width= 0.95\columnwidth]{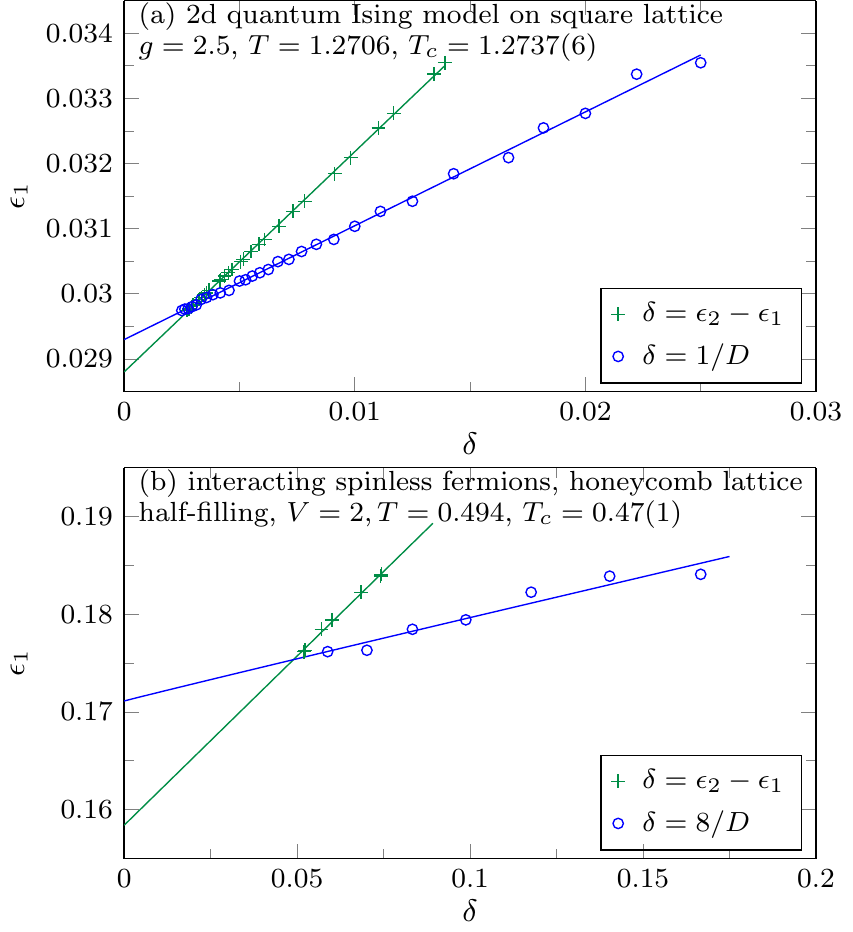}
\end{center}
  \caption{Two-dimensional quantum systems at finite temperature. PEPS description was obtained with VTNR algorithm \cite{Czarnik_2015,Czarnik_compass_2016, Czarnik_2017}. Panel (a) shows results for quantum Ising model on a square lattice with PEPS bond dimension $\chi = 5^2$ -- with square accounting for double layer of PEPS tensors. Panel (b) shows results for interacting spinless fermions on a honeycomb lattice ($\chi = 16^2$). 
  In panel (a) CTM bond dimension was $D = 40$---$400$ and $D = 50$---$140$ in panel (b).
  In both cases $\epsilon_1 (D)$ is approaching an asymptotic value nearly linearly as a function of the refinement parameter $\delta = \epsilon_2-\epsilon_1$ introduced in this article. Using $1/D$ as a refinement parameter results in less trustworthy extrapolation with an error of a few percent. The values of $T_c$, which we show for reference, were obtained from Monte Carlo simulations \cite{Hesselmann_TIsingQMC_16}. 
  } 
\label{fig:TM_PEPS}
\end{figure}      

\subsection{2d quantum states}
The PEPS corresponding to the partition function of 2d classical models -- analyzed in the previous sections  -- are given analytically. Here, we focus on 2d quantum systems, where PEPS is obtained as a result of a suitable variational procedure. We argue that our method can be applied in such case as well. In particular, we present results for a quantum Ising model on a square lattice, 
\begin{equation}
H = -\sum_{ \langle m,n \rangle } \sigma^x_{m}\sigma^x_{n} - g \sum_{m } \sigma^z_{m},
\end{equation}
where we set $g=2.5$.

We also consider a system of interacting spinless fermions on a honeycomb lattice, 
\begin{equation}
H = -\sum_{ \langle i,j \rangle } \left(c_{i}c_{j}^{\dag} + c_{j}c_{i}^{\dag}\right) + V \sum_{ \langle i,j \rangle } n_{i} n_{j},
\end{equation}
at half filling, $\langle \hat n \rangle=1/2$. We set $V=2$. Here, $c_i$ is a fermionic annihilation operator on site $i$ and $n_i=c_i^{\dag}c_i$ is a fermion number operator.  

We focus on states at finite temperature, where their PEPS description  was obtained using variational tensor network renormalization (VTNR) \cite{Czarnik_2015, Czarnik_compass_2016, Czarnik_2017}.
We show the extrapolated correlation lengths in Fig.~\ref{fig:TM_PEPS}. Again, a clean functional behavior of $\epsilon_1$ as a function of the refinement parameter $\delta = \epsilon_2-\epsilon_1$ introduced here is observed. Together with all the other results presented in this article, this allows us to conclude the superiority of our method also in such a case. 


\begin{figure*}[t!]
\begin{center}
  \includegraphics[width= 2\columnwidth]{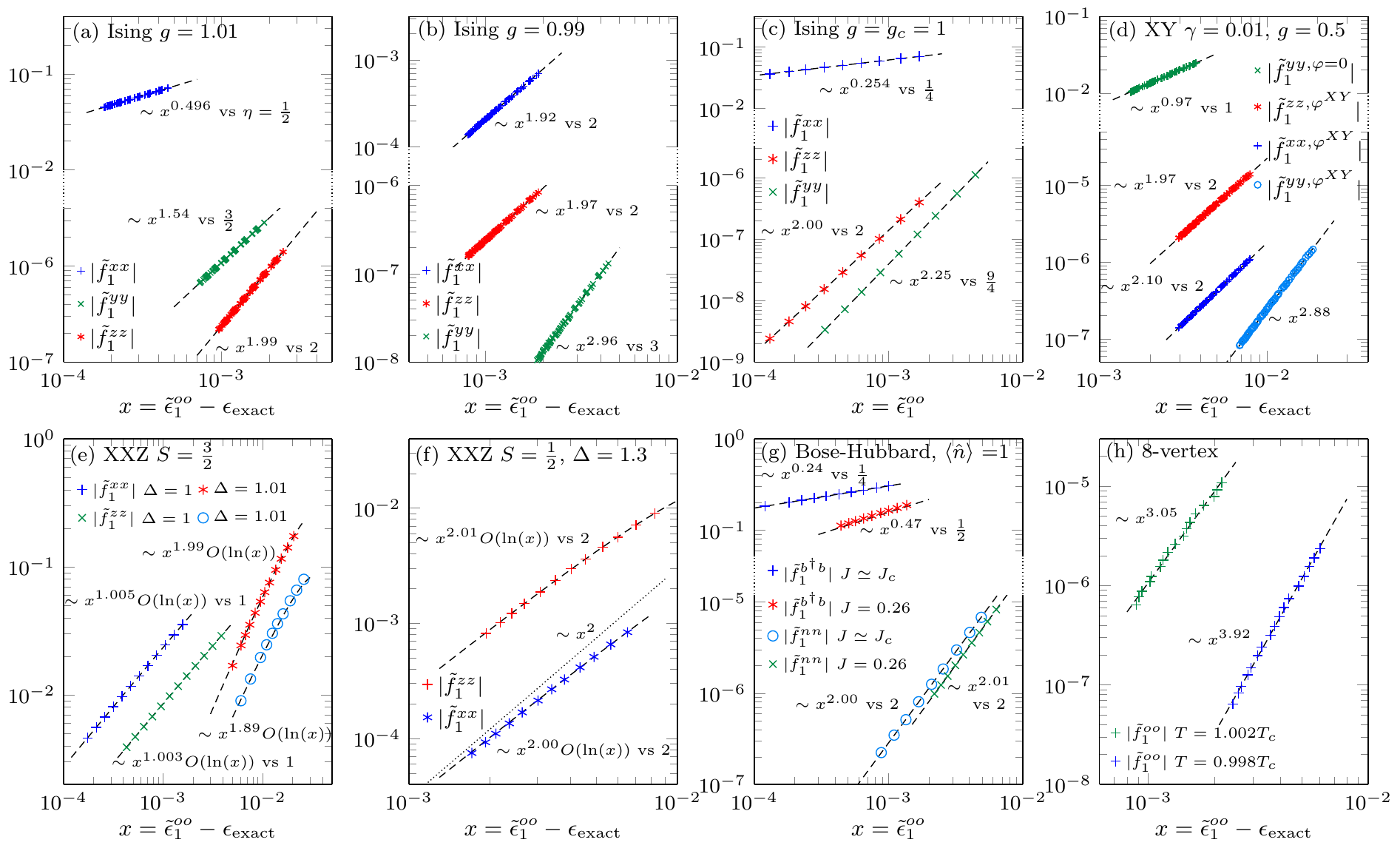}
\end{center}
  \caption{ Nonzero form factors corresponding to dominant TM eigenvalues for all the models discussed in this article. We observe that they decay (as we increase the bond dimension) as a power law with the error of the inverse correlation length, $\tilde \epsilon^{oo}_1$, extracted directly from the transfer matrix.  
  We observe that the exponent of the power law coincides with the corresponding exponent $\eta$ of the correlation function asymptotics in Eq.~\eqref{eq:C_asymptotic}. In each plot we show the fitted value of the exponent, as well as the value of $\eta$ when the analytical result is available.
  See text for details.  In each panel, the order of the legend corresponds to the respective  position of different lines. 
  Recall that  $f^{oo}_n$ is the form factor (for some $C_{oo}(R)$) corresponding to $\epsilon_n$. $\tilde \epsilon_m^{oo}$ are the eigenvalues for which the form factor is nonzero. Those nonzero form factors are marked as
  $\tilde f^{oo}_n$.
  }
   \label{fig:FF}
\end{figure*}


\section{Form factors renormalization}
\label{sec5}
In this section we focus on the behavior of the dominant form factors. We collect the data  for all the models studied in this article in Fig.~\ref{fig:FF}. For each $D$ we plot the nonzero form factors and the corresponding largest TM eigenvalue. As explained below, we plot the form factors as a function of an error of an inverse correlation length, i.e. the distance between $\tilde \epsilon_1^{oo}$ and the actual value of the correlation length inverse, $\epsilon_{\rm exact}$, for some correlation function $C_{oo}(R)$. In each panel, $D$ is increased from right to left. For the models where the exact value of the correlation length is not know analytically we use the result of our extrapolation procedure.

The main observation here is that the form factors are decreasing as a power law of an error defined above. Also, the exponent of that scaling matches the value of $\eta$ in the correlation function asymptotics in Eq.~\eqref{eq:C_asymptotic}. We base this observation on the models where $\eta$ is known analytically. Indeed, for the Ising model \cite{barouch_statistical_1971} in the paramagnetic phase, Fig.~\ref{fig:FF}(a), $\eta=1/2$, $3/2$  and $2$ for $C_{xx}(R)$, $C_{yy}(R)$ and $C_{zz}(R)$ correlators, respectively. In the ferromagnetic phase, Fig.~\ref{fig:FF}(b) we have $\eta=2$, $3$  and $2$ for the above correlators. At the critical point, Fig.~\ref{fig:FF}(c), $\eta=1/4$, $9/4$ and $2$, respectively. The above values are in very good agreement with the results of fits in Fig.~\ref{fig:FF}(a-c).

For the XY model in the incommensurate phase, Fig.~\ref{fig:FF}(d), the exponent $\eta=2$, $1$ and $2$, respectively. It is worth noting that  for $C_{yy}(R)$ the leading dependence, $\eta=1$, is obtained from the sector with complex phase $\varphi=0$, see Fig.~\ref{fig:TM_XY_ga001g05}(a). This is in agreement with the analytical result where the leading behavior is monotonic \cite{barouch_statistical_1971}. The oscillating part of $C_{yy}(R)$, with frequency $\varphi=\varphi^{XY}$, appears only with larger powers of $1/R$ in the algebraic part. Indeed, this can also be seen in the scaling of the respective form factors in Fig.~\ref{fig:FF}(d).

Before we discuss other models, we provide an argument for why such a relation is to be expected. Generally, starting with the quantum transfer matrix, the correlation function can be expressed as $C_{oo}(R) = \int d\vec k f^{oo}(\vec k) e^{-\epsilon(\vec k) R}$,
where $e^{-\epsilon(\vec k)}$ are eigenvalues of QTM, parametrized by some set of continuous parameters $\vec k$. $f^{oo}(\vec k) d\vec k$ are the corresponding form factors.
We assume that the relevant contributions in some frequency of oscillations of the correlation function can be collected as  $e^{-R/\xi} \int_{0}^{y_{\rm max}}  dy \hat f^{oo}(y) e^{-y R}$, where $y = |\epsilon(\vec k)| - \epsilon_{\rm exact}$, $\epsilon_{\rm exact} = 1/\xi$.
Integrated form factors are collected as $ \hat f^{oo}(y)  dy = \int^* d\vec k f^{oo}(\vec k) \delta(|\epsilon(\vec k)|-y)$. Here the asterisk means that we integrate over $\vec k$ contributing to given frequency of correlation function oscillations.  In order to obtain the asymptotics, $C_{oo}(R) \sim e^{-R/\xi} R^{-\eta}$, the integrated correlator $\hat f^{oo}(y)$ would have to scale as $y^{\eta-1}$ in the limit of small $y$. 

We can view the MPS TM as an approximation to the true QTM resulting from a renormalization group procedure which captures relevant degrees of freedom for the effective impurity problem along the virtual (imaginary time) direction of a true QTM \cite{Zauner_2015, Rams_2015, Bal_2016}.
This means that the dominant eigenvalue of the TM contributing to some $C_{oo}(R)$, $\tilde \epsilon^{oo}_1$, should represent contributions from the range of the smallest $y \in [0,y_1]$, with  $\tilde \epsilon^{oo}_1-\epsilon_{\rm exact}  \sim y_1$. The corresponding form factor is then obtained by averaging over the same range, $\tilde f^{oo}_1 \simeq \int_{0}^{y_1} \hat f^{oo}(y) dy$. Collecting those results, together with the expectation that $\hat f^{oo}(y) \sim y^{\eta-1}$,  we obtain 
\begin{equation}
\tilde f^{oo}_1 \sim (\tilde \epsilon^{oo}_1-\epsilon_{\rm exact})^{\eta}. 
\label{eq:fx}
\end{equation}
The above argument is rather qualitative and should be understood as representing general intuition of what renormalization of the (virtual) degrees of freedom of QTM looks like in the numerical procedure leading to given MPS approximation with finite bond dimension.
It is, however, hard to expect that a more formal derivation is possible, because the numerical algorithm is directly targeting variational energy as a figure of merit. As a result, it is not directly related to the full TM.
Nevertheless, the results of this section give strong support to the above argument.

We can further test it in the other models considered in this article. The XXZ spin chain provides an interesting example. The usual correlation function asymptotics is modified by logarithmic corrections at the isotropic critical point. It is true for spin-$\frac12$ and spin-$\frac32$ models. The theoretical prediction reads
$C_{xx}(R) = C_{zz}(R) \sim \frac{\sqrt{\ln (b R)}}{R^\eta}$ with $\eta = 1$ \cite{Giamarchi_XXZ12_1989, Singh_XXZ12_1989, Affleck_1989}. This relation is supported by numerical studies of finite systems \cite{Hallberg_XXZ12_1995, XXZ32_Hallberg_1996}. We further corroborate this in the Appendix \ref{sec:app} using results of our iDMRG simulations. Those logarithmic corrections, which are stronger in the spin-$\frac32$ case, should also manifest themselves in the scaling of the form factors. Similarly as above, in this case we expect to recover the following relation 
\begin{equation}
\tilde f^{oo}_1 \sim (\tilde \epsilon^{oo}_1-\epsilon_{\rm exact})^{\eta} \sqrt{\ln\left(\frac{d}{\tilde \epsilon^{oo}_1-\epsilon_{\rm exact}} \right)} \ .
\label{eq:fx_log}
\end{equation} 
For XXZ spin-$\frac32$ model at the critical point, we test it in Fig.~\ref{fig:FF}(e) and obtain excellent agreement with the above prediction.
The situation, however, becomes less clear in the gapped phase. For $\Delta=1.01$ shown in the log-log plot, Fig.~\ref{fig:FF}(e), the scaling clearly deviates from the straight line, i.e. pure power law.
If we assume that the vicinity of the critical point is still influencing the scaling (at least for this range of $D$) and allow for logarithmic corrections also in this case, then the data become consistent with $\eta\approx2$, for both $C_{xx}(R)$ and $C_{zz}(R)$. 

Such claims can be supported by analyzing the XXZ spin-$\frac12$ model where the situation is similar, although the effects of logarithmic corrections are weaker. 
In the massive phase, $\Delta=1.3$, shown in the log-log plot, Fig.~\ref{fig:FF}(f), the scaling deviates again from the clear power law. If we however assume that the logarithmic corrections are still relevant here, we should include them in the fit using Eq.~(\ref{eq:fx_log}). This allows to recover the exponent $\eta \approx 2$ for both $C_{xx}(R)$ and $C_{zz}(R)$. This is in very good agreement with the analytical results, $\eta = 2$ \cite{Dugave_XXZmassive_2015, Dugave_XXZmassive_2016}. The dotted line $\sim x^2$ in the plot serves as a guide for the eye.

In the Bose-Hubbard model with unit filling the situation seems to be much simpler. In  Fig.~\ref{fig:FF}(g) we analyze the form factors for $\langle  b^\dagger_0  b_R \rangle$ and $ \langle  n_0  n_R \rangle$ and the power-law fits allow us to obtain the values of $\eta$ very close to the ones predicted  theoretically \cite{Haldane_QF_1981, Giamarchi_BH_1992, Giamarchi_BH_1997, Larkin_JA_1997, Monien_BH_2000}. For $J=0.304$, very close to the critical point, we obtain $\eta = 2.00$ and $\eta = 0.24$ respectively. The exact values are $2$ and $\frac14$.
In the Mott insulator phase, $J=0.26<J_c$, we have $\eta = 2.01$ and $\eta = 0.47$, respectively, which should be compared with the expected $2$ and $\frac12$.

Finally, in the 8-vertex model for $J=0.2$, $J'=J''=0.1$ simulated here, we analyze the form factors corresponding to the order parameter. From the power-law fits, see Fig.~\ref{fig:FF}(h), we obtain $\eta = 3.05$ and $\eta = 3.92$ for ferroelectric ($T = 0.998T_c$) and disordered ($T=1.002 T_c$) phases respectively. 

To conclude this section, a few remarks are in order.  We should note that the power-law fits in Fig.~\ref{fig:FF} are susceptible to the value of the inverse of the true correlation length $\epsilon_{\rm exact}$, which is especially relevant when the exact value is not known. As a result, errors of up to a couple of percent might be expected here. Nevertheless, we can contrast the indirect approach presented here with fitting the asymptotics of the correlation function directly. As we illustrate in Appendix \ref{sec:app}, the latter can give substandard results away from the critical point within MPS simulation. As shown there, even in the simple Ising model in the ferromagnetic phase, the values of both $\xi$ and $\eta$ cannot be extracted with high accuracy from fitting Eq.~\eqref{eq:C_asymptotic} directly.  The indirect method discussed in this section reduces the error by an order of magnitude.

Finally, the data shown for all the models in this section are consistent and well described by the scaling in Eqs.~\eqref{eq:fx} and \eqref{eq:fx_log}. Additionally, they are in good agreement with the analytical values of $\eta$ (when available). This allows us to gain deeper understanding about the information encoded in the MPS TM and its relation with the true QTM \cite{Zauner_2015, Rams_2015,Bal_2016}. It also allows us to expect good accuracy also for the models where the value of $\eta$ is not otherwise known. 

\section{Conclusion}
\label{sec6}
The main message of this article is that some care has to be taken when extracting long-distance properties from MPS and CTM simulations. The extrapolation procedure introduced here allows to extract the correlation length with much better accuracy than the widely used method based on the bond dimension. For one-dimensional systems, in the vicinity of the critical point, it can be used in parallel with the finite entanglement scaling to corroborate the results of the latter. We note that, for systems in two dimensions, one has to combine the two approaches. In that case our method can be used as a prerequisite to control CTM contraction error and set up finite correlation length scaling \cite{Rader_FCLS_2018, Corboz_FCLS_2018, Czarnik_2018}.

We test our procedure, among others, in the gapped phase of the one-dimensional Bose-Hubbard model with unit filling. This allows us to fit the scaling form of the correlation function in the vicinity of the BKT critical point and extract, among others, the position of the critical point with high accuracy.

We also discuss how the algebraic part of the correlation function asymptotic is directly encoded in the scaling of the form factors. This provides a new tool for calculating this quantity within MPS simulations, which, especially away from criticality, might be much better than fitting the asymptotics directly.

\section{Acknowledgement}
We thank Philippe Corboz and Jacek Dziarmaga for insightful discussions. We acknowledge support by National Science Center, Poland under Projects No. 2016/23/D/ST3/00384 (MMR) and No. 2016/23/B/ST3/00830 (PC). This research is supported in part by the U.S. Department of Energy through J. Robert Oppenheimer fellowship (LC).

\appendix
\section{Direct fitting of the correlation function asymptotics}
\label{sec:app}

\begin{figure} [t]
\begin{center}
  \includegraphics[width= 0.95\columnwidth]{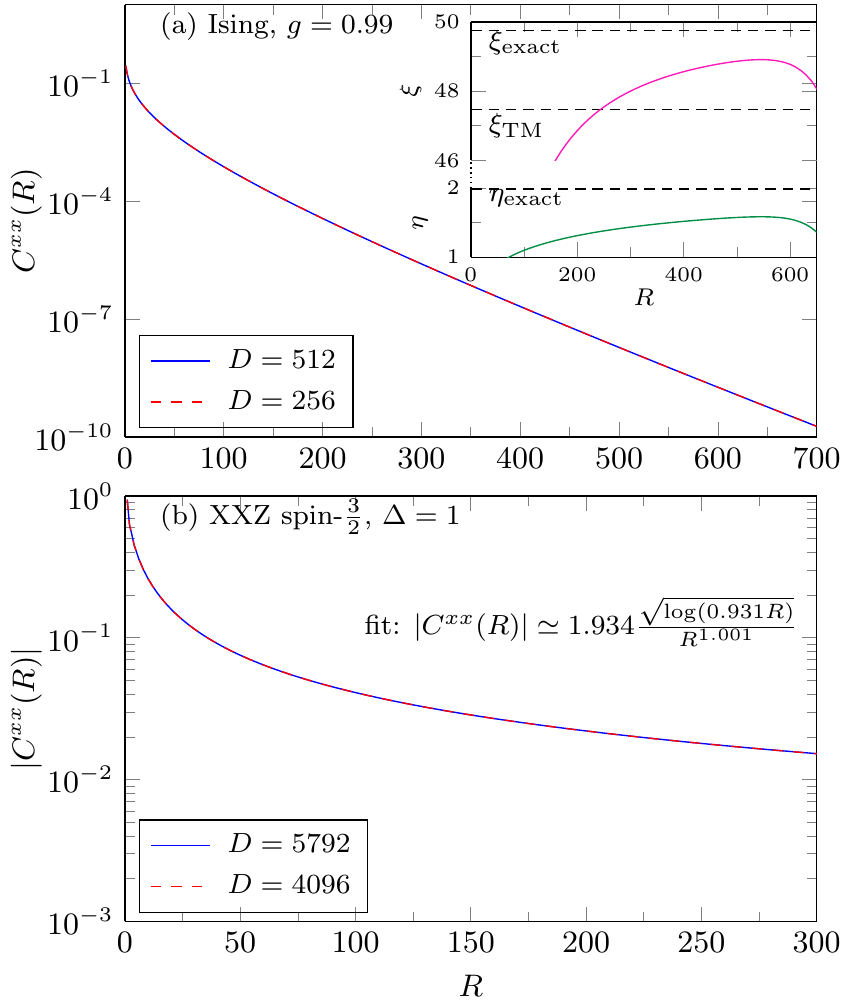}
\end{center}
  \caption{ In panel (a) we show an example of direct fitting of the correlation function asymptotics from Eq.~\eqref{eq:C_asymptotic} for the Ising model in the ferromagnetic phase, $g=0.99$.
  The results of the fit $\ln C_{xx}(r) \simeq a + r/\xi - \eta \ln(r)$ for windows of $r\in [R -25,R +25]$ are shown in the inset. Extracting the actual correlation function asymptotic is substandard in this case.
  For comparison,  in panel (b) we show the correlation function for the critical antiferromagnetic Heisenberg spin-$\frac32$ chain, which can be very well fitted with the theoretical scaling prediction. The fit was done for $R\in[100,300]$.
  }
   \label{fig:ap1}
\end{figure}

In this appendix, we illustrate problems related to fitting the correlation function asymptotics directly, which provides further evidence of superiority of the extrapolation scheme proposed in this article. 
In Fig.~\ref{fig:ap1}(a), we show $C_{xx}(R)$  in the ferromagnetic phase of the Ising model. We check that the results are converged in the bond dimension. We fit the asymptotic form in Eq.~\eqref{eq:C_asymptotic} for different ranges of the distance $R$ to that data.

We expect three regimes of $R$. 
For relatively short distances $1\ll R \ll \xi$ the behavior of the correlation function is still strongly influenced by the vicinity of the critical point, for which $C_{xx}(R) \sim R^{-\eta_1}$, with $\eta_1 = 1/4$ in our example.
In the other extreme limit, $R \gg 1/\delta$, where $\delta=\tilde \epsilon^{xx}_2 - \tilde \epsilon^{xx}_1$ is the measure of deviation from the continuous spectrum,
we end up with purely exponential behavior, $e^{-R \tilde \epsilon^{xx}_1}$, i.e. the one in Eq.~\eqref{eq:C_MPS}. This is an ultimate consequence of finite bond dimension used in the numerical simulations. Finally, let us see if one can recover the exact asymptotic, $e^{-R/\xi} R^{-\eta}$, in the intermediate regime $\xi \ll R \ll 1/\delta$. The combination of required scale separation, numerical precision and limitations imposed by finite bond dimension makes this interesting limit very hard to attain in practice, especially when $\eta \neq \eta_1$ (here, $\eta=2$).

\begin{figure} [t]
\begin{center}
  \includegraphics[width= 0.95\columnwidth]{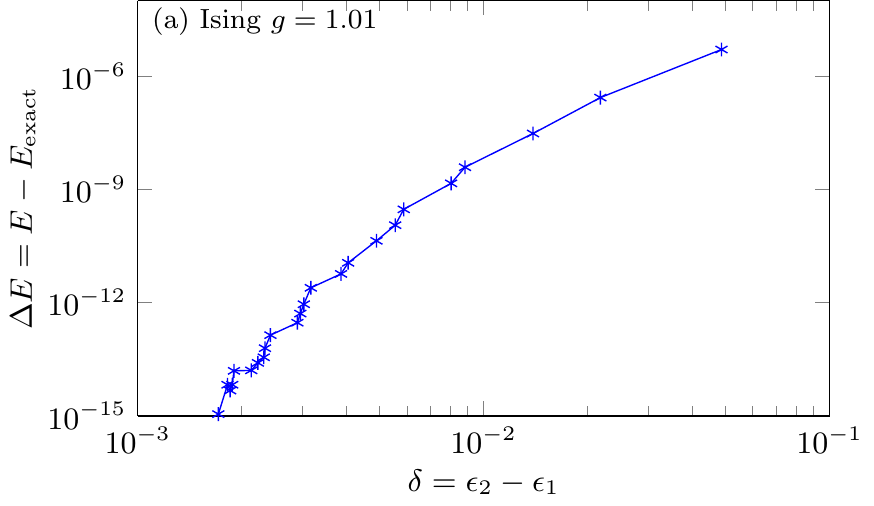}
  \includegraphics[width= 0.95\columnwidth]{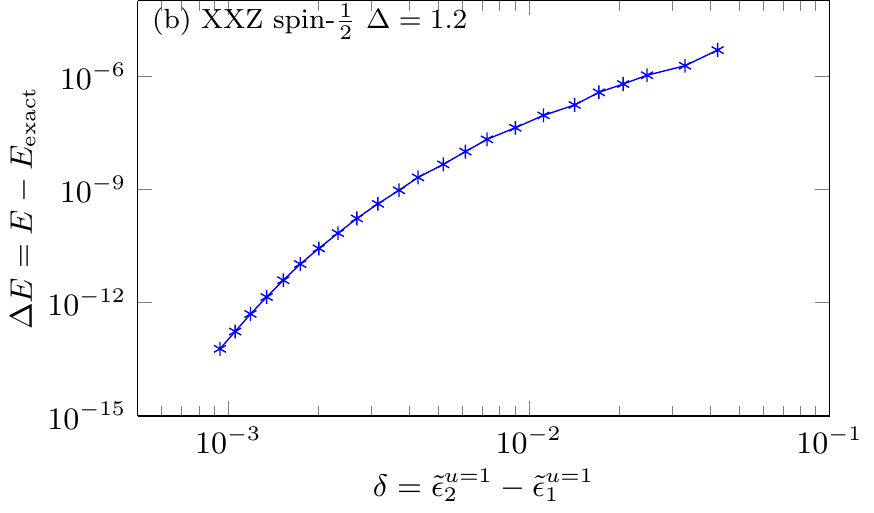}
\end{center}
  \caption{ The refinement parameter $\delta$ introduced in this article is not suited for extrapolation of local quantities, such as energy. This is illustrated using the example of (a) Ising model and (b) XXZ spin-$\frac12$ model, where we show the error of ground state energy $\Delta E$ converged for different $D$. In panel (a) $D=8$---$100$ and in (b) $D=32$---$1448$. This implies that the methods which do work well for the extrapolation of the energy would not work for the correlation length.
  }
   \label{fig:ext_E}
\end{figure}

A smooth transition between such three limits can indeed be recognized in Fig.~\ref{fig:ap1}(a). In our example $\xi = 49.7\ldots$ and $1/\delta \approx 700$ ($D=512$), as can be read from Fig.~\ref{fig:TM_Ising}(d). It is worth observing that the largest values of locally fitted correlation length, even though they are larger than the value given by the largest TM eigenvalue, are still almost $2\%$ away from the exact one. Additionally, they strongly depend on the window of $R$'s used in the fit. As a result, extracting it in this way is not very reliable. For comparison, our extrapolation procedure gives a result that is an order of magnitude more accurate, see Fig.~\ref{fig:TM_Ising}(d).  Similarly $\eta \approx 1.5$, which is the largest local estimate in Fig.~\ref{fig:ap1}(a) is far from being precise. A significantly better estimate is obtained from scaling of the form factors in Fig.~\ref{fig:FF}(b). We stress here that while the discussion above is quite general, the illustrative example of the Ising model is well known not to be challenging for MPS-based methods. As such, we could expect even more severe problems for more demanding systems.
 
Paradoxically, the situation at the critical point is much simpler, even though such points are generally harder to simulate with MPS. Above all, there is no physical length scale $\xi$ in this case. As such, we can expect to recover the exact asymptotics for $1 \ll R \ll 1/\delta$, with finite $D$ effects becoming relevant on larger distances only. This, in principle, makes direct extraction of the correlation function asymptotics much more straightforward in this case (there might still be problems related, e.g., with spontaneous symmetry breaking resulting from finite $D$, or other effects of inefficient description of critical points with MPS).

As an illustrative example, in Fig.~\ref{fig:ap1}(b) we show $C_{xx}(R)$ for the isotropic antiferromagnetic Heisenberg spin-$\frac32$ chain. The difference between the results for two bond dimensions shown there, as well as difference from $C_{zz}(R)$, are well below $10^{-5}$.
In this case the scale $1/\delta \approx 3000$, as can be read from Fig.~\ref{fig:TM_XXZ32}(a). Correlation function asymptotics was theoretically predicted \cite{Affleck_1989, Giamarchi_XXZ12_1989, Singh_XXZ12_1989} as $ (-1)^R a \frac{\sqrt{\ln(b R)}}{R^\eta}$, with $\eta = 1$. 
From the fit to the region $R\in[100,300]$, which is order of magnitude smaller then $1/\delta$, we obtain $a=1.9344(81)$, $b = 0.931(19)$ and $\eta=1.00148(41)$ with sum of squared errors $\mathrm{SSE}= 3 \cdot 10^{-10}$, in very good agreement with the prediction. This further corroborates and improves upon the relatively old  verification \cite{XXZ32_Hallberg_1996} obtained in the DMRG study of finite systems of up to $60$ spins.
 
 The above example, apart from corroborating older results, clearly shows that extracting the correlation function asymptotics from MPS simulations at the critical point is a viable   method of obtaining the exponent $\eta$, unlike for the system away from criticality. Indeed, even logarithmic correction is clearly recovered in the example and has to be taken into account in the fit.  Likewise, we should note that obtaining $\eta$ in the critical systems from the fits to the form factors, as in Sec~\ref{sec5} and Fig.~\ref{fig:FF}(c,e,f) seems to yield comparable precision of the results. The latter method, however, proves to be superior away from the critical point.

\section{Extrapolation of the local quantities}
\label{sec:app2}

The refinement parameter $\delta$ introduced in this article proves to be well suited for extrapolation of nonlocal quantities such as the correlation length.
The natural question is if it could be used for extrapolation of local observables, such as energy per site or order parameter as well. To resolve this question, in Fig.~\ref{fig:ext_E} we show the error of the ground state energy $\Delta E$ as a function of $\delta$ for the Ising and XXZ spin-$\frac12$ models. The relation $\Delta E(\delta)$ is not particularly smooth and this shows that such approach is not suited for local quantities. The results presented in Fig. \ref{fig:ext_E} should be compared with those in Figs.~\ref{fig:TM_Ising}(a) and \ref{fig:TM_XXZ12}(a) for the Ising and XXZ spin-$\frac12$ models respectively. This, however, allows us to argue that one should not expect a smooth relation between the error of the correlation length and the error of the ground state energy.

There are well-established approaches to extrapolate the energy in MPS \cite{legeza_1996, white_2005, Leblanc_2015, ehlers_2017, Hubig_extrapolation_2017}. They are based on the truncation error or the energy variance as a refinement parameter for such fits. The argument above, however, implies that they should not be used to extrapolate the correlation length.


%

\end{document}